\documentclass[twocolumn]{aastex62}
\usepackage{placeins} % remove when draft is ready.
\usepackage{natbib}
\usepackage{apjfonts} 
\usepackage{xcolor}
\usepackage{multirow}
\newcommand{\msun}{M$_{\sun}$}

\newcommand{\Lsun}{L$_{\sun}$}

\newcommand{\Teff}{$T_{eff}$}
\newcommand{\lgTeff}{$log_{10}(T_{eff}/\rm K)$}

\newcommand{\e}[2]{$\rm #1 \times 10^{#2}$}
\newcommand{\powten}[1]{$\rm 10^{#1}$}

\accepted{2019 October 2}
\submitjournal{ApJ}

\shorttitle{Convective Overshoot in SN IIP Progenitors}
\shortauthors{Wagle et al.}
\begin{document}

%\title{Convective Overshoot In Type II-P Supernova Progenitors: Surface Composition, Blue Loops And Their Explodability}
\title{Type IIP Supernova Progenitors and their explodability I: Convective Overshoot, Blue Loops and Surface Composition}

\correspondingauthor{Gururaj A. Wagle}
\email{guru.w84@gmail.com}

\author[0000-0002-3356-5855]{Gururaj A. Wagle}
\affiliation{Homi Bhabha Centre for Science Education - Tata Institute of Fundamental Research, Mankhurd, Mumbai 400088, India}

\author[0000-0003-2404-0018]{Alak Ray}
\affiliation{Homi Bhabha Centre for Science Education - Tata Institute of Fundamental Research, Mankhurd, Mumbai 400088, India}

\author{Ajay Dev}
\altaffiliation{Participants in the National Initiative on Undergraduate Science (NIUS)\\ program
at HBCSE (TIFR)}
\affiliation{National Institute of Science Education and Research Bhubaneswar, 752050, India}

\author{Adarsh Raghu}
\altaffiliation{Participants in the National Initiative on Undergraduate Science (NIUS)\\ program
at HBCSE (TIFR)}
\affiliation{Indian Institute of Science Education and Research Kolkata, 741246, India}


\begin{abstract}

\deleted{We present the evolution of massive star progenitors of supernovae of type IIP taking the example of the nearby and well-studied SN 2013ej. The evolution from the pre-main sequence phase upto their core-collapse stage is studied with the help of Modules for Experiments in Stellar Astrophysics (MESA) code. It shows significant variations of many surface and internal properties with the extent of core and envelope convective overshoot.} \added{We present the evolution of massive star progenitors of supernovae of type IIP. We take the example of the nearby and well-studied SN 2013ej. We explore how convective overshoot affects the stellar structure, surface abundances, and effective temperature of massive stars, using the Modules for Experiments in Stellar Astrophysics (MESA).} In particular models with moderate overshoot \added{($f$ = 0.02 to 0.031)} show the presence of blue loops in the Hertzsprung-Russell diagram with a red to blue \added{[\lgTeff \ from $< 3.6$  to $> 4.0$] excursion} and transition back to red, \deleted{after core helium ignition}\added{during core helium burning phase}. \deleted{unlike those with with higher or lower overshoot which} \added{Models with overshoot outside this range of $f$ values} kept the star in the red supergiant state throughout the post helium ignition phases. \deleted{Surface CNO abundance and the blue to red supergiant population ratio in a steady state are also affected by convective overshoot.} \added{The surface CNO abundance shows enrichment post-main-sequence and again around the time when helium is exhausted in core. These evolutionary changes in surface CNO abundance are indistinguishable in the currently available observations due to large observational uncertainties. However, these observations may distinguish between the ratios of surface nitrogen to oxygen at different evolutionary stages of the star.}
%\added{The Blue to Red Supergiant population ratio ranges between 0.5 to 2.0 for our simulations that exhibit blue loops, comparable to the observed ratio of 0.6 for the Large Magellanic Cloud which has similar metallicity as our simulations.} 
\deleted{We compare our models with the observed properties of Large Magellanic Cloud supergiants due to their close resemblance of metallicity. We also evaluate the effects of convective overshoot on the 'compactness factor' of the stellar core which may control the probability of actual explosion as opposed to the collapse into a black hole. The compactness parameters resulting from moderate overshoot make it easier for the star to explode as a SN.} \added{We also compare the effects of convective overshoot on various parameters related to likelihood of explosion of a star as opposed to collapse to a black hole. These parameters are the compactness parameter, M$_4$, and $\mu_4$. Combination $\mu_4 \times $M$_4$, and $\mu_4$ have similar variation with $f$ and both peak out at $f$ = 0.032. We find that all of our 13 \msun \ models are likely to explode.} 

\end{abstract}



\section{Introduction} \label{sec:intro}

Type IIP supernovae \added{are core-collapse supernovae which} have red supergiant stars (RSG) as progenitors that retain large hydrogen envelopes at the time of their explosion. \added{This is indicated by} pre-explosion images of the sites of core collapse supernovae \citep{Smartt:2009ab,Smartt:2004aa}\deleted{ indicate this}. However, the progenitor star of SN 1987A retained a substantial hydrogen envelope but exploded from a blue supergiant (BSG) stage \citep{Arnett:1989aa}. \deleted{Even so, }This star was previously a RSG and was on a "blue loop" when it reached the penultimate BSG stage before exploding \citep{Woosley:1988ab,Woosley:1988aa}. \deleted{For observational reasons, the identification of the progenitors with their supernovae can be carried out for relatively nearby supernovae (at $d \leq 30 \rm \; Mpc$).} 

\deleted{The structure of the progenitor as well as the development of the explosion that results from it differ significantly depending upon the mass, metallicity and chemical composition, internal mixing, stellar rotation and magnetic fields, mass loss in stellar wind, binarity, etc. While this reflects a myriad of (often unknown) factors that may affect the "mapping" of the stellar properties on to the properties of the observable supernovae that may result from it,}\added{A fundamental problem of stellar evolution and explosion entails how the properties of the progenitor star determine the characteristics of the supernova resulting from it. Some of these properties are the mass, metallicity and chemical composition, internal mixing, stellar rotation and magnetic fields, mass loss in stellar wind, binarity, etc. An observational limitation here is that the unambiguous identification of the progenitors of the supernovae can be carried out only for relatively nearby supernovae (at $d \leq 30 \rm \; Mpc$). Therefore,} few carefully selected and well-observed supernovae may offer significant insights. 

From this perspective, we study here and in the companion Paper II (Wagle \& Ray, submitted to ApJ) \added{and Paper III (Wagle et al., in preparation)}, the evolution of a few specific cases of some of the variables mentioned above. \added{We do this }with a view to simulating models of explosions seen in e.g. SN 2013ej that occurred in M74 at a distance of 9.0$^{+0.4}_{-0.6}$ Mpc \citep{Dhungana:2016aa}. For these studies, we take the overall metallicity in the vicinity of the progenitor the same as in the case of SN 2013ej \citep{Chakraborti:2016aa} that is Z = 0.006. \added{This metallicity was determined in turn} from the metallicity of a nearby H II region number 197 of \citet{Cedres:2012aa} in M74. Our main objective here is to assess the dependence of stellar properties at various evolutionary stages upon the convective overshoot and resultant chemical mixing. We use the publicly available code Modules for Experiments in Stellar Astrophysics \citep[MESA,][]{Paxton:2011aa,Paxton:2013aa,Paxton:2015aa,Paxton:2018aa} to investigate both pre-supernova stars' properties as well as their explodability. We simplify the other input physics and assume our stars to be non-rotating and non-magnetized. We primarily investigate the case of an isolated (single) 13 \msun \ star discussed in detail in Paper II Chandra X-ray studies point to a progenitor Zero Age Main Sequence (ZAMS) mass of 13.7 \msun \ \citep{Chakraborti:2016aa}. We also assume a standard mass loss (Dutch scheme) implementation in MESA described in the Methods section below. 

\added{Our aim here is to study how convective overshoot affects three key aspects of the evolution of supergiant stars, namely, 1) how these stars undergo blue loops in the Hertzsprung-Russell diagram (HRD) after reaching a RSG stage, 2) how their surface CNO abundances evolve with stellar age for different overshoot extent, and 3) how the internal structure of the deep core in the star may be affected by overshoot.  This region of the star may determine whether the star explodes as a supernova or collapses to a black hole. We also compare the modelled properties to relevant observational data on supergiant stars in the Large Magellanic Cloud.}
\added{In our future paper III, we shall compute blue to red supergiant ratios from our model simulations and compare them with the observed data for the specific metallicity environment.}

In section \ref{sec:background}, we briefly review the conceptual background and \deleted{past results }\added{previous work in the field} relevant to our study. In section \ref{sec:methods}, we describe the computational methods and details of stellar evolution using the MESA code. In section \ref{sec:obs}, we outline the method of selection of observational data of supergiants to compare with our model calculations in the Hertzsprung-Russell diagram (HRD) and surface composition evolution. In section \ref{sec:results}, we describe how convective overshoot affects a whole range of stellar properties. This includes whether the final core structure at the pre-supernova (pre-SN) stage is conducive to explosion in a supernova resulting in a neutron star versus collapse into a black hole, and the relation of the blue loops to the external convection zone. In section \ref{sec:conclusion}, we discuss our results and summarize our conclusions.

\section{Effects Convective Overshoot on Supergiant Stars: A Background} \label{sec:background}

Almost all one dimensional stellar evolution codes treat convection in the mixing length theory (MLT) formulated by \citet{Bohm-Vitense:1958aa}. MLT however is not able to describe the   boundary region between the stably stratified region that sits adjacent to the convective region, as in its formulation the flow of matter stops at the boundary (v=0). However, the inertia of the convective flow would in fact penetrate the stably stratified region beyond \citep{Viallet:2015aa,Arnett:2015aa} leading to mixing in the region beyond the already mixed convective region. Usually the overshoot of the convective layers refer to the extra mixing of chemical elements beyond the convective boundaries defined in the MLT picture. \citet{Viallet:2015aa} delineate three regimes of overshooting  depending upon the importance of radiative effects. The MESA code implements the simplest of these regimes where overshooting does not affect the thermal structure of the layers beyond the convective boundary, but only leads to chemical mixing. That is, here the effect of radiation is dominant and the {\it P\'eclet number, Pe,} which is a ratio of the time scale for radiative transport of heat to the time scale for advective transport of heat {\it is of the order of unity or smaller}. In the overshooting zone where a thermal diffusion dominated situation persists ($Pe \leq 1$), turbulent eddies of all scales are only able to mix the composition, without changing the entropy structure significantly.

\subsection{Blue Loops} \label{subsec:blue_loop}

The question of occurrence (or non-occurrence) of blue loops in the HRD was considered by \citet{Stothers:1991aa}. Downward overshooting of the star's external envelope when the convective envelope reaches its greatest depth, causes the hydrogen abundance discontinuity to dig deeper into the interior. This may lead to a much closer approach of the hydrogen-burning shell to the hydrogen abundance discontinuity during the core helium depletion when the hydrogen shell burns through \added{outward}\deleted{ in mass coordinate}. They point out that because their models near the bottom of the RSG branch are only marginally stable, small inward displacements of the outer convection zone lead to convective overshoot by a factor $D_{ov}/H_P$ = 0.3 \deleted{(equivalent to $f =0.3$ for MESA)}\added{\citep[in step overshoot formalism, see, e.g.][]{Pedersen:2018aa} and}\deleted{ to} promote blue loops. \deleted{However, their choice of the overshoot factor were perhaps on the high side as seen in this paper (see our Results section). }\added{\citet{Schaller:1992aa} found that the step overshoot parameter is slightly reduced with newer opacities than \citet{Stothers:1991aa}. \citet{Schaller:1992aa} fitted their models to observed terminal age main sequence of several stellar clusters and associations to find a best fit model that uses $D_{ov}/H_P$ = 0.2 over the range from 1.25--25 \msun .}

\added{\citet{Chin:1991aa} investigated evolutionary sequence of stellar models with convective overshoot for the stellar mass range $3 - 30 \; \rm M_{\odot}$, but with much higher (supersolar) initial metallicity $Z = 0.021 - 0.044$. They found that convective overshooting during the main sequence phase leads to a shortening of the blue loop that forms when the helium in the core is being depleted, while overshooting from the helium burning core has the opposite effect. They also found that larger initial abundance of metals tends to shorten the blue loop. At the same time, blue loops can form when downward convective overshooting of the outer convection zone is allowed to occur \citep{Stothers:1991aa}. }

\added{
\citet{Tang:2014aa} computed evolutionary tracks of massive stars \added{using the PARSEC code} at low metallicity (Z = 0.001 and Z = 0.004) and compared with observational data of color-magnitude diagrams (CMDs) in metal-poor irregular star-forming dwarf galaxies. \added{They treated step overshooting at the base of the convective envelope only, with $D_{ov}/H_P$ = 0.7.} While their CMDs are reproduced reasonably well, the predicted blue loop in their calculations is not hot enough with \added{this} extent of overshooting. \added{They recommended a higher overshoot $D_{ov}/H_P$ = 2-4 to explain the extended blue-loops observed in low metallicity galaxies.} %\citet{Jin:2015aa} examine how the mixing length parameter and the overshooting affect the properties of convective cores and convective envelopes in massive stars. Convective core overshooting during the main sequence phase, makes the star enter the RSG phase earlier than the case without core overshooting and that envelope overshooting can facilitate the formation of the blue loop. 
}

\added{\citet{Saio:2013aa} explored the evolution of massive stars using Geneva code model calculations with solar metallicity (Z = 0.014 and 0.020) and initial stellar masses 8--50 \msun \ . They compared models with and without rotational mixing. They note that the rotational mixing increases the size of the helium core. This results in faster red-ward transition from BSG post-MS, and hence, a higher mass loss rate due to increased luminosity. The blue loop is triggered earlier than the model that does not include rotation. However, similar results can be achieved in a non-rotating model if a more extensive overshoot in the core is used. They find that a (step) overshoot higher than $D_{ov}/H_P \sim$ 0.4 would result in an extensive blue-loop for a 20 \msun \ non-rotating star.}

\citet{Paxton:2013aa} (MESA paper II) investigated the \added{HRD and the} convective core growth in low and intermediate mass non-rotating stars during their main sequence and core helium burning stages\deleted{ (as well as their HRD)} and explored various mixing options. They recommend that when the overshoot mixing is treated with the option of exponential decay \added{(refer to section \ref{sec:methods}),} the overshoot factor should be $f_{ov} \sim 0.016$\deleted{ in contrast to the larger convective overshoot factors $0.1 < f_{ov} < 0.16$ that multiplies the pressure scale height $H_P$ that determines the overshoot zone as fully mixed }. \added{Their recommendation was based on stellar evolution models of \citet{Herwig:2000aa}, which used exponential decay formalism for convection overshoot. \citet{Herwig:2000aa} could reproduce models of \citet{Schaller:1992aa} with $f_{ov} \simeq$ 0.016.} Our overshoot factors $f$ used below include the small $f \sim 0.016$ value but spans also a set of slightly larger values to investigate the presence of blue loops\deleted{, though it is not as large as in the literature quoted in \citet{Paxton:2013aa}}.

\added{
\citet{Deng:1996aa} presented evolutionary models for massive stars computed with turbulent diffusion \citep{Deng:1996ab} for chemical mixing during overshoot. They analyzed the HRD with observational data on, e.g. the LMC \citep{Fitzpatrick:1990aa}.% They noted that despite many successes of stellar evolution theory, there were difficulties in simultaneously matching the model properties with those of observed properties of luminous stars in the HRD and those of the progenitor of the SN 1987A in the LMC. They identified the nature of convection and its associated mixing as the most important factors, but mass loss from the star in stellar wind as another important consideration. 
The new formulation of \citet{Deng:1996aa} was able to deal with convective overshoot and semiconvection at the same time within the Padua stellar evolution code. This allowed them to simultaneously recover results of the wider main sequence band, high luminosity and long lifetimes of massive stars found thus far in standard overshoot models and the extended blue loops typical of models run with semiconvection models. %[We note that modern stellar evolution codes like MESA (e.g. version-r10398) allow options that can simultaneously turn on overshoot and semiconvection as we have undertaken in this paper and in our Paper-II. In this paper our focus has been to investigate the different extents of convective overshoot (with a fixed extent of semiconvection) within the "exponential diffusion coefficient model" of the overshoot.] 
Among the features that the model calculations of \citet{Deng:1996aa} reproduced were the existence of many stars in the so-called "blue gap", the "ledge" invoked by \citet{Fitzpatrick:1990aa} \citep[see also][]{Tuchman:1990aa,Ray:1991aa,Rathnasree:1992aa}; the number frequency distribution across the HRD for specific
metallicity environments, e.g. surface abundances at the terminal stage (of their calculation); blue to red supergiant ratios \citep[see also][]{Langer:1995aa}; etc. In addition they discussed the life times of the central H and He-burning phases and internal composition structure of three prototype sequence of stars of masses  6, 20 and 60 \msun with standard overshoot. Nevertheless, %as already mentioned 
%it remains a challenge to reproduce 
the reproduction of the characteristics of the blue supergiant progenitor of SN 1987A remained a challenge, despite the successes of stellar evolution theory.
}

\deleted{The blue to RSG ratio in an ensemble of massive stars such as a galaxy or a set of young clusters as a function of metallicity Z has been argued to be a steeply rising function of increasing metallicity. %\added{The blue to RSG ratio has been argued to be a steeply rising function of increasing metallicity in an ensemble of massive stars such as a galaxy or a set of young clusters \citep[see ][and references therein]{Langer:1995aa}}. 
The ratio is sensitive to mass loss, convection and other mixing processes. The ratio and its Z-dependence obtained from a number of clusters in the Milky Way galaxy and the Magellanic Clouds have been revisited by \citet{Eggenberger:2002aa} who also summarize the previous work in this area. They give a normalized relation of the B/R (to the solar neighborhood value) in terms of metallicity as: 
${(B/R) / (B/R)_{\odot}} \sim 0.05 \times e^{{3 Z/Z_{\odot}}}$.
Since convective overshoot is an important ingredient of mixing in both the nuclear burning core or its edge as well as in the stellar envelope surrounding it, the B/R ratio 
%\added{in a restricted region of the Hertzsprung Russell diagram (HRD)} 
may be an important diagnostic of the extent of convective overshoot as well. }

\subsection{Surface Chemical Abundance}

\added{The external convection zone (when present) dredges out the CNO products from the hydrogen shell burning layers to the surface. These elements' abundance and their mutual ratio may then be different from the primordial gas from which the star formed. Similarly, if the external convection zone thoroughly mixes elements from the hydrogen burning shell at an earlier stage, the effect of later enhancement from a subsequent dredge-up may be masked. The presence of external convection zone in the star (or the lack thereof in intermediate models) may therefore affect the surface abundance of these elements. The overshoot factor $f$ is a controlling factor for the extent of the convection zone. \citet{Meynet:2015aa} studied the tracks in the HRD for massive stars with metallicity Z = 0.014 with the help of the Geneva stellar evolution code. They investigated the positions of the pre-SN progenitor in HRD and the structure of the stars for various mass-loss rates during the RSG phase for two different initial rotation velocities. They also discussed the surface trace element ratios e.g. (N/C) and (N/O) during the RSG phase and compare with galactic RSGs. %Enhanced mass-loss rates from these solar metallicity stars produce significant changes in the populations of BSGs, YSGs, and RSGs and extended blue loops. They predict that a majority of blue (yellow) supergiants under these circumstances are post-RSG objects. However, as far as properties of presupernova stars are concerned, yellow pre-SN stars require an enhanced mass loss rate while the RSG pre-SN progenitors are well fitted by standard mass loss rate. They point out the properties of post-RSG progenitors of several supernovae, among which there are a few type IIP/IIL SNe, e.g.  SN 2008cn, SN 2009hd, SN 2009kr etc., but these progenitors are only mildly post-RSG stars in terms of $T_{eff}$. Production of a BSG progenitor like that of SN 1987A, at metallicities of about Z= 0.006 and with ZAMS masses between 15 and 25 \msun remains a challenge; these  \citet{Meynet:2015aa} say may provide clues on the physics of rotational mixing and of mass loss.
They found that these ratios can be reproduced by the non-rotating models or models with low initial rotation. }

\deleted{
\citet{Davies:2019aa} present a method of inferring the mass of the progenitor of a type IIP SN using the observed abundance ratios at a very early stage after the explosion of the SN. During this early phase (within a few days) while the photosphere is very hot, optical spectra show ionized carbon, nitrogen and oxygen as also hydrogen and helium. They argue that predictions from stellar evolution calculations for RSGs for the terminal surface [C/N] ratio is correlated with the initial mass of the progenitor star. The use of very early spectra of the supernova according to them facilitates the estimation of the pre-explosion carbon, nitrogen, and oxygen abundances, since the photosphere being hot, high ionization species of such elements allow the spectra to be dominated by these elements. While very early time spectra of supernovae have been somewhat rare so far (e.g. SN 2013fs = iPTF13dqy), where O \textsc{vi}, O \textsc{v}, O \textsc{iv} and N \textsc{v} lines have been detected at $T_{in} = 48-58 kK$ \citep{Yaron:2017aa} or SN 2016esw \citep[typically at 0.4-0.6 d after explosion,][]{de-Jaeger:2018aa} modeling such lines as O \textsc{v} and O \textsc{vi} have not been very reliable. Moreover, very early spectra are often heavily dominated by high continuum emission and line characteristics may be more difficult to determine.
}

As already noted, in a blue loop, a BSG has evolved from a {\it previous RSG} phase, as opposed to the BSG evolved directly from the main sequence phase. There are methods to distinguish BSG stars on the (first) transit to the RSG branch as against those which have already been a RSG during its previous evolution based on surface composition abundance \citep{Maeder:2014aa}. \citet{Saio:2013aa} have tried to devise a diagnostic based on stellar pulsation to discriminate BSGs with these different evolutionary histories \citep[see also][]{Bowman:2019aa}. \deleted{Using model calculations with solar metallicities and initial stellar masses in 8--50 \msun \ range, they}\citet{Saio:2013aa} found that {\it radial} pulsations are excited in the BSGs only if they have previously been RSGs. In addition they found that for a given surface temperature, models that evolve after the RSG phase have many more excitations of {\it non-radial} pulsations than those which are evolving towards the RSG phase. \added{They however note that not all of their 14 \msun \ models on the blue-loop excite pulsation. The pulsations are triggered only during the red-ward transition while in the loop.  For lower mass stars, the pulsations are not triggered even if they are on the blue loop, as their L/M ratio is too small to excite pulsations.} In their first study \citet{Saio:2013aa} found that the computed surface abundance ratios N/C and N/O in stars with rotational mixing after having evolved through a RSG stage seem high compared to observations of the supergiant stars Rigel and Deneb, even though their pulsational properties point to their RSG past. \added{They found that the N/C ratio was about a factor of 10-15 times higher for 14 \msun \ model, while it was 20-30 times higher for 25 \msun \ model than what was observed for the two stars. Similarly, the N/O ratio was about a factor of 3-4 higher for 14 \msun \ model and a factor of 6-9 higher for 25 \msun \ model.}  This puzzle was later resolved by \citet{Georgy:2014aa} who showed that the use of ``Ledoux'' criterion for convection (as opposed to the ``Schwarzschild'' criterion) significantly improved the agreement with both observed surface composition in addition to pulsational periods. \added{With a 25 \msun \ model sequence using ``Ledoux" criterion, they showed that both the surface abundances ratios come down to within a factor of 2-3.5 of observations.} \deleted{We show below that the properties of supergiant stars are affected by the nature of convective overshoot.}

\subsection{Explodability }

\added{The location of the lower boundary of carbon burning convective shells outside the oxygen rich core is an important factor \citep{Sukhbold:2014aa} in determining the compactness parameter at core bounce\footnote{$\xi_{M}$ is a dimensionless variable that allows a prediction of post-bounce dynamics and the possible evolution towards a black hole formation \citep{Oconnor:2011aa}. It is instructive to see how the internal structure of the core represented by this parameter evolves as a star goes through the advanced stages of nuclear evolution in its core as in \citet{Sukhbold:2014aa}. The mass M = 2.5 \msun \ was chosen since this is the mass scale for black hole formation. On the other hand, \citet{Ugliano:2012aa,Ertl:2016aa} have suggested that a different fiducial mass 1.75 \msun \ may be a better discriminant of the explosion characteristics. As noted by \citet{Sukhbold:2018aa}, linked convective shells give a structure with smaller compactness parameter $\xi_M$ and thus favor SN explosions rather than BH formation.} during the core-collapse}
\begin{equation}
\xi_M = \frac{M/M_{\odot}}{R(M_{baryonic} = M)/1000 km}\Big |_{t-bounce}
\end{equation}
\added{It is well known that the \added{success or failure of explosion}\deleted{ possibility versus collapse to a black-hole} is related to the strength and evolution of the accretion after stellar core-bounce \citep[see e.g.][and references therein]{Ugliano:2012aa}. Initial investigations of the condition for explodability of presupernova stellar models by \citet{Oconnor:2011aa} indicated that explosion is the likely outcome of core collapse for progenitors with $\xi_{2.5} \leq 0.45$ if the nuclear equation of state (EOS) of the dense core matter is favorable. In contrast they considered that progenitors with $\xi_{2.5} > 0.45$ would most likely lead to black hole formation as a result of stellar collapse. Note however that the critical neutrino heating efficiency required for explosion plotted against the compactness factor $\xi_{2.5}$ for various EOS for each progenitor star model showed considerable scatter in their Fig 8.} 

\added{While the expansion of the stalled shock of the initially exploding star is retarded by the ram pressure of the infalling stellar-core matter overlying the stalled shock, the shock expansion\footnote{\added{\citet{Colgate:1989aa} pointed out that as inelastic collisions behind the shock front convert the relative kinetic energy of moving matter to heat, they lead to the increase of the internal energy of the flow. This rendered the flow subsonic just behind the strong shock. \citet{Colgate:1971aa} showed that when the initial driving pressure due to the core elastic bounce subsumes, the resulting rarefaction wave would soon be able to catch up with the shock front, which would then weaken. The presence of a hot, high-entropy bubble behind the shock due to deposition of neutrino energy in the hot bubble \citep{Mayle:1991aa,Wilson:1993aa} rescues the situation referred to above. The shock would ultimately re-invigorate only due to the presence of this hot, high-entropy bubble that kept pushing against the matter outside the proto-NS (PNS) and made the shock resume.%s its outward expansion. %\citet{1971ApJ...163..221C} showed that the matter adjacent to the proto neutron star immediately after the bounce cools so rapidly by neutrino emission that re-implosion occurs and a rarefaction wave catches up with the matter that was previously moving out with the bounce shock.
}} (``revival'') is assisted by neutrino energy deposition behind the stalled shock \citep{Bethe:1985aa,Ertl:2016aa} beyond a critical neutrino luminosity \citep{Ray:1987aa,Burrows:1993aa}. At the same time the mass accretion rate on the proto neutron star, the source of the external ram pressure can be related to $\xi_{2.5}$ which is higher for denser stellar cores. \citet{Ertl:2016aa} proposed a two parameter criterion which would in combination predict the explosion behavior for neutrino driven supernova. \citet{Ertl:2016aa} calculated the outcome of explosion dynamics based on 1-D hydro calculations and schematic models of the neutrino emission from a hot and high density PNS using methodology of \citet{Ugliano:2012aa}. As inputs to these calculations, they used pre-SN stellar structure models based on both Kepler code \citep[e.g.][]{Woosley:2002aa} as well as Japanese stellar evolution code \citep[see e.g. ][]{Nomoto:2006aa}. They computed more than 600 simulations and obtained more than 97\% success rate of explodability prediction. These two parameters are $M_4$ and $\mu_4$ where $M_4$ is the enclosed mass (normalized to solar mass) where the dimensionless entropy per nucleon reaches $s = S/k_B = 4$ and 
\begin{eqnarray} \label{eqn:mu4}
\mu_4 & = & \frac{(\delta m/ M_{\odot})}{(\delta r/1000 km)}|_{s=4} \nonumber \\
		   & = & \frac{(M_4 +\delta m/M_{\odot})-M_4}{[r(M_4 +\delta m/M_{\odot}) - r(s=4)]/1000 km }
\end{eqnarray}
is the normalized mass derivative at this location (where $S/k_B =4$). Here $\delta$m = 0.3 \msun .
\citet{Ertl:2016aa} found correlations between $\mu_4$ and $\mu_4 \times \rm M_4$ from the explosion simulations for quantities such as ZAMS mass, explosion energies,etc. for both successful as well as failed explosions (see their figures 11 \& 12). These correlated parameters for the successful versus failed explosion cases had two distinct regions in the relevant dimensional parameter space.}

%%%% Figure %%%%%%
\begin{figure}[htb!]
\includegraphics[width=0.49\textwidth]{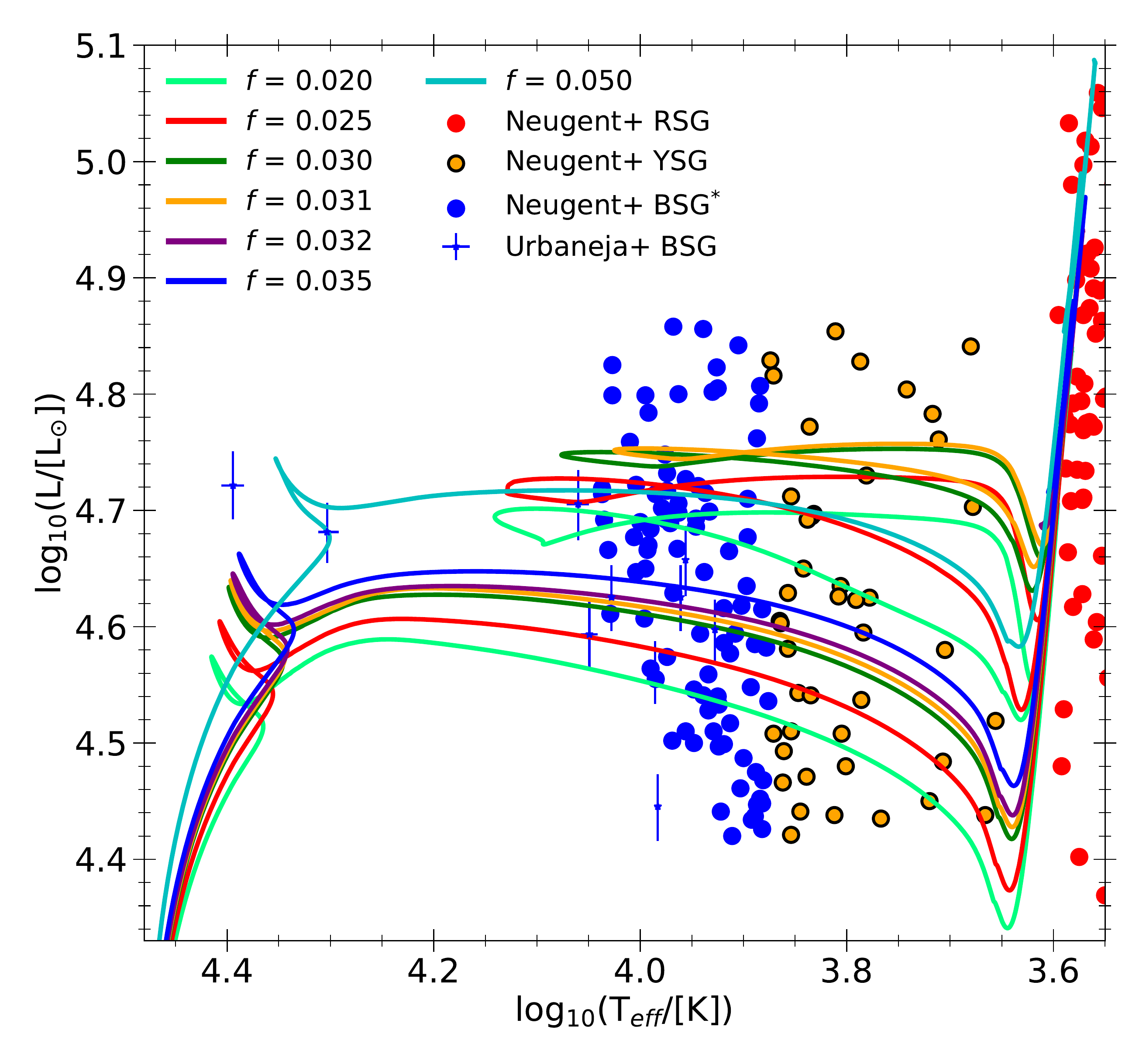}
\caption{The HRD comparing the models with M$_{ZAMS} $= 13 \msun \  and Z = 0.006 for different values of overshoot parameter, $f$ with $f_0$ held constant at 0.005.  The models with $0.020 \le f \le 0.032$ go through a blue loop. The archival observations for Large Magellanic Cloud (LMC) stars are overlaid on the graph. \added{Details of the selection criteria and uncertainties for these data are given in section \ref{sec:obs_HR}. [$^*$Note that the YSG catalog from \citet{Neugent:2012aa} also includes stars with \lgTeff \ $\ge$ 3.875 and upto 4.0 (circles marked in blue). %\citeauthor{Neugent:2012aa} note that they have been overly generous by including stars in this range as YSG candidates in their observations. 
We consider these stars as BSG candidates.]}
\label{fig:13M_HR}}
\end{figure}
%%%%%%%%%%%%%%%

\section{Method of Simulations} \label{sec:methods}

We use MESA version r-10398 to model a star such as the progenitor star of a type IIP supernova SN 2013ej. We use a 13 \msun \ ZAMS mass with initial metallicity Z = 0.006 for the simulations. We evolve the star from the pre-Main-Sequence (pre-MS) stage through core collapse (CC) in MESA. %A detailed list of physical parameters used for our MESA simulations of the stellar evolutionary models is provided in appendix A of paper II. 
The MESA inlists files used for our simulations presented here are made available publicly at MESA market place\footnote{cococubed.asu.edu/mesa$\_$market/} (subject to acceptance of the manuscript.)\deleted{ We are submitting a sample inlist file for referee's consideration)}. The MESA inlists parameters used in our simulations are discussed in detail in Methods section of Paper II, and in relevant MESA instrument papers. All simulations presented in this paper use the exact same mass and temporal resolution controls, wind scheme for mass loss, isotopes network and reaction rates, opacities and equation of state options, mixing length theory ($\alpha_{MLT}$), semiconvection, stopping criterion, etc. as the 13 \msun \ model 3 with 79 isotopes network presented in paper II. The only parameter that we have experimented with in this paper is the overshoot parameter $f$.

In the absence of 3-D hydrodynamic treatment of convection, the hydrodynamic instabilities at convective boundaries, namely convective overshooting, must be accounted for through parametric modeling. MESA uses exponential diffusive overshooting beyond the convective boundaries defined by the Schwarzchild criterion \citep{Herwig:2000aa}. MESA calculates the diffusive coefficient for overshoot mixing
\begin{equation} \label{eqn:D_ov}
D_{OV} = D_{conv,0} \ exp\left(\frac{-2(r-r_0)}{f\lambda_{P,0}}\right)
\end{equation}
Here $D_{conv,0}$ is the MLT derived diffusion coefficient calculated at a user specified location near the convective boundary inside the convective region. As the \added{diffusive coefficient} $D_{conv}$ in MLT goes to zero at the convective boundary (r$_{CB}$), \deleted{the code calculates this coefficient} \added{MESA uses a numerical work around by calculating the convective diffusive coefficient $D_{conv,0}$} at a location $r_0$ \added{close enough to the boundary set by MLT}\deleted{inside the convective region} \added{, but allows chemical mixing even beyond the boundary. In this overshoot region, standard MLT would not permit mixing due to vanishing convective velocity.} \deleted{such that} \added{The location} $r_0 = r_{CB} - f_0 \lambda_{P,CB}$, \deleted{where $f_0$ is an user defined parameter, and} \added{(}where $\lambda_{P,CB}$ is the pressure scale height at the Schwarzchild boundary $r_{CB}$,\added{) is where the new (non-zero) diffusion coefficient is defined. The quantity} $\lambda_{P,0}$ \added{in equation \ref{eqn:D_ov} is then} the pressure scale height at the location $r_0$. \deleted{This is a numerical trick to calculate the convective diffusion coefficient close enough to the boundary.} %\added{By doing so MESA effectively uses $(f+f_0)$ instead of just $f$ in equation \ref{eqn:D_ov} \citep[see also][]{Pedersen:2018aa}. Users can vary values of both $f$ and $f_0$ in MESA.}
This is strictly chemical mixing and the thermal structure does not change. The switch from convective mixing to chemical mixing happens at $r_0$. \added{The parameter $f_0$ defines the position and therefore the normalization of the diffusion coefficient (i.e, $\rm D_{conv,0}$) given in equation \ref{eqn:D_ov}. On the other hand, $f$ determines the decay width (in radial coordinates) of the diffusion coefficient, and therefore the extent of radial mixing.} More details of the effects of overshooting are discussed in \citet{Pedersen:2018aa,Viallet:2015aa}. If $f_0$ is small we remain closer to the convective boundary. For all of the simulations presented in this paper we have chosen a single value of \deleted{overshoot} parameter $f_0$ = 0.005, as we find no dependence of the existence of blue loops on this parameter. \deleted{The overshoot parameter $f$ has been varied between 0.02 and 0.035 in steps of 0.005. We have also tried a higher value of $f$ = 0.05, and some intermediate values of $f$ = 0.03 and 0.035, where the blue loop appears towards lower end of $f$ but disappears at $f$ = 0.032 (see Table \ref{tab:br_ratios} for all of the $f$ values).} \added{We explored a range of values for overshoot parameter $f$ starting at a low value of $f$ = 0.010 to a value of $f$ = 0.050, which spans a factor of 5. We observed that the blue loop behavior is absent both below $f$ = 0.020 and above $f$ = 0.031. Hence, we present here only representative models with selected $f$ values of 0.020, 0.025, 0.030, 0.031, 0.032, 0.035 and 0.050.}

%%%% Figure %%%%%%
\begin{figure*}[htb!]
\gridline{\fig{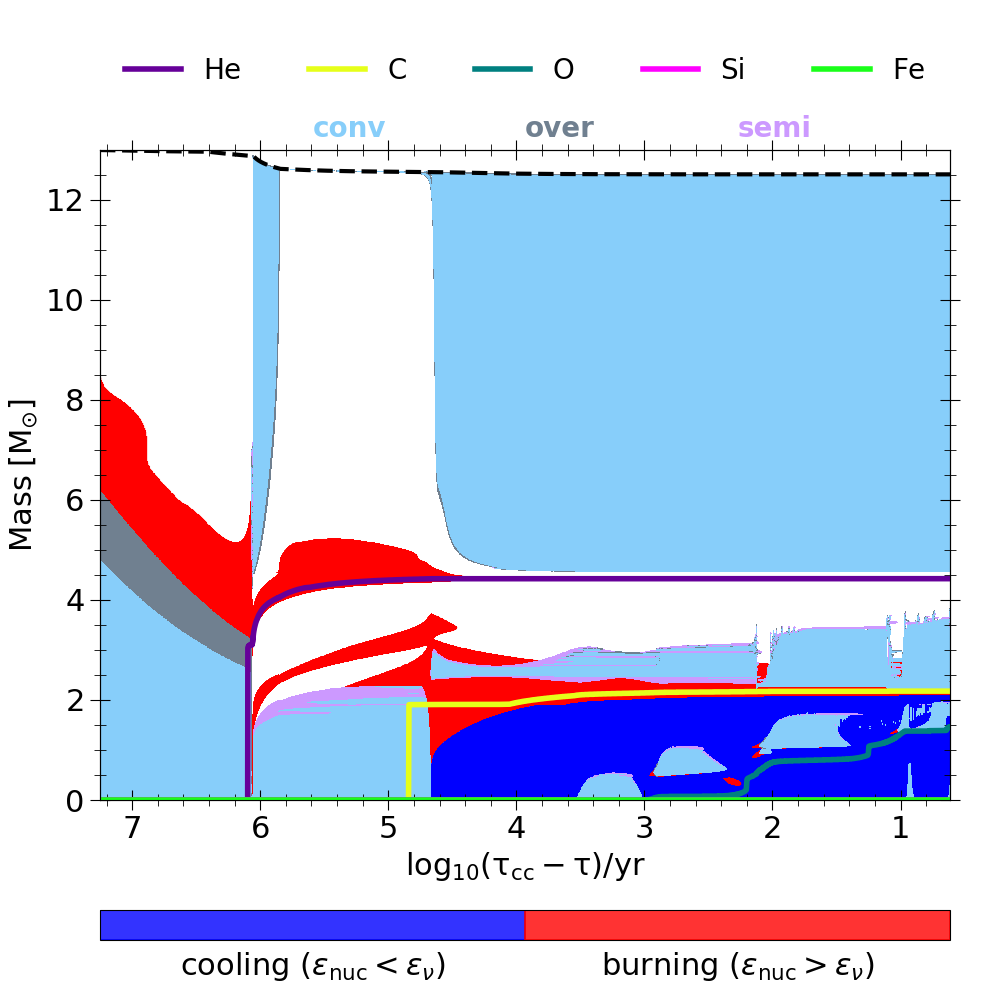}{0.45\textwidth}{(a)}
			  	\fig{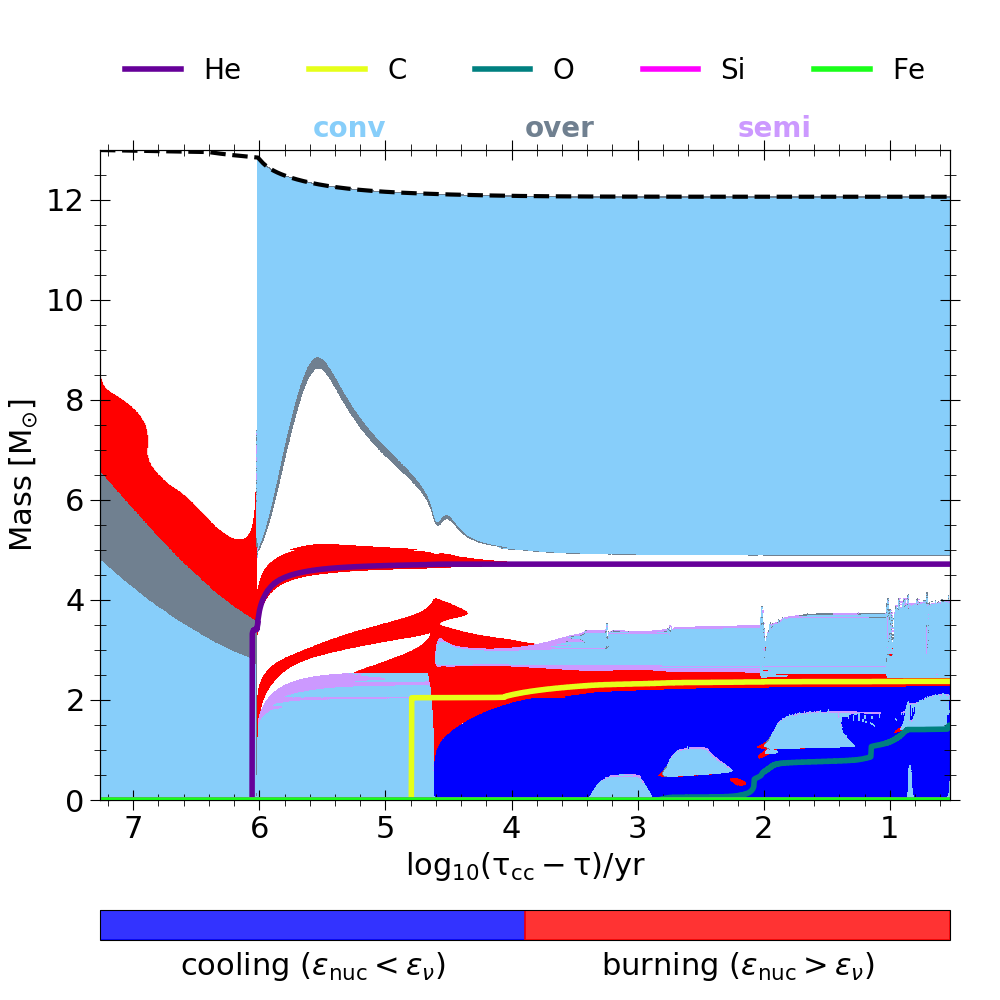}{0.45\textwidth}{(b)}}
\gridline{\fig{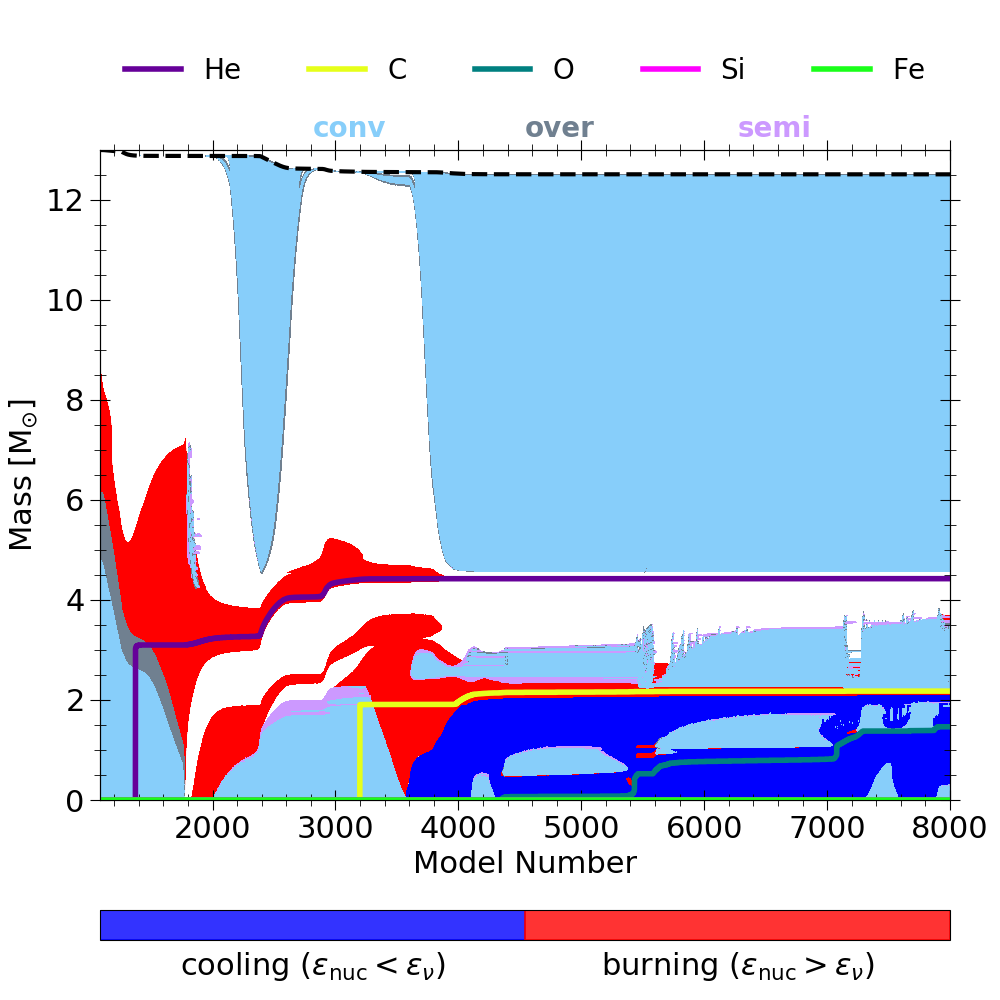}{0.45\textwidth}{(c)}
			  	\fig{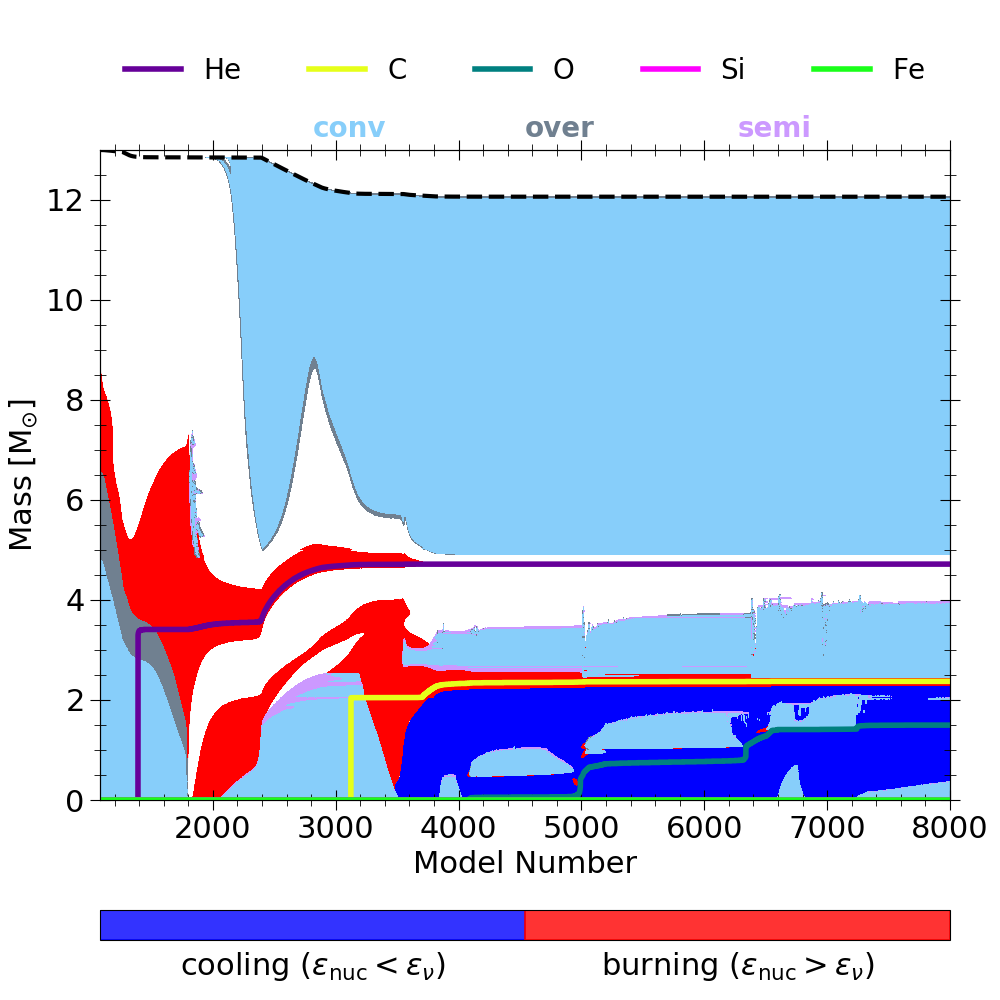}{0.45\textwidth}{(d)}}
\caption{Comparison of the Stellar structure (Kippenhahn diagram) for two of the M$_{ZAMS}$= 13 \msun \  and Z = 0.006 models with values of overshoot parameter $f$ = 0.025 (left panels) and 0.032 (right panels) is shown. The top panels show the plots as a function of time before CC, and the bottom panels show the plots as a function of MESA model numbers, for the same range of data. The star on the left with $f$=0.025 exhibits a blue loop, while the one on the right with $f$=0.032 is at the threshold where the blue loop disappears (ref. figure \ref{fig:13M_HR}). The other parameter $f_0$ related to overshoot is held constant at 0.005. \added{The core boundaries for He, C, O, Si, and Fe reported by MESA are plotted in the diagram.  The default definition of a core boundary in MESA is where the mass fraction of previous (dominant) isotope decreases below a threshold value of 0.01 and the same for the current isotope is above 0.1. The total star mass is also shown in the diagram as a dashed black curve at the top.}
\label{fig:13M_kipp}}
\end{figure*}
%%%%%%%%%%%%%%

\section{Archival Observations} \label{sec:obs}

\subsection{Selection of LMC supergiants for H-R Diagram} \label{sec:obs_HR}

The stars plotted in Fig. \ref{fig:13M_HR} include RSG and YSG \added{candidates} \deleted{taken}\added{selected} from \citet{Neugent:2012aa} and BSG \added{candidates} \deleted{taken}\added{selected} from \citet{Urbaneja:2017aa}. \added{\citet{Neugent:2012aa} selected YSG and RSG candidates using the color dependent magnitude cutoff's as described in their section 2 (also see their Figure 2). They used stellar models to define a cutoff for K magnitude as a function of J$-$K magnitude for the YSG candidates down to 12 \msun \ between a $T_{eff}$ range of 4800-7500 K (3.875 $>$ log$_{10}$\Teff \ $>$ 3.68). However, they included candidates upto 10,000K (log$_{10}$\Teff = 4) in their YSG catalog. The cooler stars below the J$-$K limit were labeled as RSG. They further eliminated the foreground candidates using their radial velocities.} \deleted{Out of the supergiants reported by \citet{Neugent:2012aa} those} \added{We selected 323 such YSG+RSG candidates} which belong to LMC (labeled Category 1\footnote{stars with radial velocity higher than 200 $kms^{-1}$. Since the radial velocity separation was very large between the LMC (centered around 278 $kms^{-1}$) and Milky Way (centered around 0 $kms^{-1}$), LMC membership could be easily proved \citep{Neugent:2012aa}} \deleted{in that paper}) and had the spectral type reported \deleted{were selected}\added{\added{in the catalog}}. \added{We then restricted the RSG candidates} \deleted{Among these supergiants the ones labeled as RSG were further restricted} within the luminosity range of $4.2 < log(L/L_\odot) < 5.1$ \deleted{and log$_{10}$\Teff > 3.55}, and the \deleted{ones labeled as }YSG \added{candidates} \deleted{were restricted} to the luminosity range of $4.42 < log(L/L_\odot) < 4.8$. \added{Our choice of these luminosity ranges are motivated by the span of the theoretical tracks of evolution in our 13 \msun \ simulations.} This finally resulted in the selection of 117 YSGs and 46 RSG \added{candidates. As noted above, the YSG candidates in \citet{Neugent:2012aa} include stars hotter than 7500 K. We marked such candidates in our selected YSG sample separately as BSG to distinguish from YSG candidates in our Fig. \ref{fig:13M_HR}. \citet{Neugent:2012aa} claim that the uncertainties in their data are comparable to the size of the points in their figure 6. Thus, the uncertainties in these data are smaller than the size of the circles in our Fig. \ref{fig:13M_HR}.}

Out of the 90 BSGs reported in \citet{Urbaneja:2017aa}, \added{we selected} 10 BSG \added{candidates} which were in the same luminosity range \added{used above to restrict} \deleted{as} the YSG \added{candidates} \deleted{above} and \added{had} \lgTeff \ < 4.5\deleted{ were selected}. The luminosity \added{for these BSG candidates} was calculated from the given $A_V$ and bolometric corrections values \added{in} \citet{Urbaneja:2017aa} for these stars using equation \ref{eqn:lum} below:
\begin{equation} \label{eqn:lum}
log(L/L_\odot)=-0.4((m_v+5-A_V-5log(d))+BC-M_{bol\odot})
\end{equation}
Distance to LMC was taken as $d$ = 49.97 \added{$\pm$ 0.19 (statistical) $\pm$ 1.11 (systematic)} kpc \citep[\added{distance modulus, $\mu_d$ = 18.493 $\pm$ 0.008(statistical) $\pm$ 0.047 (systematic) mag; ref.}][]{Pietrzynski:2013aa}. The errors were propagated for \added{the uncertainties in luminosity} \deleted{\citet{Urbaneja:2017aa} data to find the uncertainties} using the following equation:
\begin{equation} \label{eqn:lum_err}
\frac{\Delta L}{L_{\odot}}=-  0.4 \;  ln(10) \frac{L}{L_{\odot}} \sqrt{\left(\Delta{A_V}\right)^2+\left(\frac{5 \Delta{d}}{d \times ln(10)}\right)^2+\left(\Delta{BC}\right)^2}
\end{equation}
\deleted{Fig. \ref{fig:13M_HR} also includes stars chosen from Table 1 of \citet{Hunter:2007aa} that are within 12--14 \msun \ range.  All of these data are provided in the machine-readable format online. The contents of the online table are specified in Table \ref{tab:obs_HR}.}\added{These error bars are plotted in Fig. \ref{fig:13M_HR}.}

\added{The observational data plotted in Fig. \ref{fig:13M_HR} is available in machine readable format online. The contents of the online table are given in Table \ref{tab:obs_HR}}

%%%% Table %%%%%%
\begin{deluxetable*}{clll}
\tablecaption{Contents of Observational Data of LMC Supergiants for HRD \label{tab:obs_HR}}
\tablehead{
\colhead{Column No.} & \colhead{Label} & \colhead{Units} & \colhead{Explanation} \\
}
\startdata
1 & Star & - & Star Name \\
2 & 2MASS & - & 2MASS identifier \\
3 & Vmag & mag & Visual Magnitude(apparent) \\
4 & U-B & mag & U-B color index \\
5 & B-V & mag & B-V color index \\
6 & V-R & mag & V-R color index \\
7 & SpType & - & Spectral Type \\
8 & log(Teff) & $log(K)$ &  log of Effective Temperature (in Kelvin) \\
9 & log(L) & $log(L_{\sun})$ & log of Luminosity (in terms of solar Luminosity) \\
10 & e\_log(Teff) & $log(K)$ & Lower uncertainty in log of Effective Temperature \\
11 & E\_log(Teff) & $log(K)$ & Upper uncertainty in log of Effective Temperature \\
12 & e\_log(L) & $log(L_{\sun})$ & Lower uncertainty in log of Luminosity \\
13 & E\_log(L) & $log(L_{\sun})$ & Upper uncertainty in log of Luminosity \\
14 & AV & mag &  Extinction Coefficient \\
15 & e\_AV & mag & Upper Uncertainty in Extinction Coefficient \\
16 & E\_AV & mag & Lower Uncertainty in Extinction Coefficient \\
17 & BC & mag & Bolometric Correction \\
18 & e\_BC & mag & (Symmetrical) Uncertainty in Bolometric Correction \\
19 & StarType & - & Classification of star (acc. to source) \\
20 & Reference & - & Source of data, (a) for data from \citep{Urbaneja:2017aa}, (b) for data from  \\
 &  &  &  \citep{Neugent:2012aa} 
\enddata
\tablecomments{Only a portion of this table is shown here to demonstrate its form and content. Machine-readable version of the full table is available.}
\end{deluxetable*}
%%%%%%%%%%%%%%%%

\subsection{Surface C,N,O Abundances} \label{sec:obs_surf_abund}

Table 9 of \cite{Hunter:2007aa} gives the photospheric abundances of stars (from UVES, FLAMES at VLT) which were determined using TLUSTY non-LTE model atmospheres code.\deleted{ (only stars within mass range of 12-14 $M\odot$ were included in our selection). Table 4 of \cite{Evans:2004aa} gives the surface abundances for stars (from UVES, FLAMES at VLT), of which only one star falls in the range of Fig. \ref{fig:13M_NC_NO}. The CNO abundances are given in Table \ref{tab:CNO_abund}, which is also available online in the machine-readable format.
}
\added{\citet{Maeder:2014aa} confirm some CNO products mixing for supergiant stars in the LMC but they argue that the large scatter in the observational points is due to the insufficient accuracy of the data for testing the various models of stars. They reanalyzed a fraction of the early B-type stars in the VLT-FLAMES data set of \citet{Hunter:2008aa,Hunter:2009aa} using their own spectral line formation computations, and gave arguments for why part of the observational data should not be used for this comparison. Their verification of ionization equilibria and abundance determination (see their Table 1) showed that rather few of their results made a good match of abundances of He \textsc{i}, He \textsc{ii}, C, N, O, Mg \textsc{i} and Si \textsc{iii}, \textsc{iv}. Among their stars listed under LMC (9 in all) several are showing double lined spectra (spectroscopic binaries) or are Be stars. Among the other stars, only one, N11-095 have all of He \textsc{i}, C \textsc{ii}, N \textsc{ii}, O \textsc{ii} abundances well matched. Unfortunately this star is too blue ($T_{eff} = 26,800 K$) to be of consequence for our comparison in the $L - T_{eff}$ space. Note that the published VLT-FLAMES observational results so far pertain only to blue part of the HRD.} \added{We find that the typical error bars derived from \citet{Hunter:2007aa} range between 16-50\% for X(CNO). The errors are propagated from the observed uncertainty in each of the C, N, and O surface abundances to the uncertainty in X(CNO) using a standard deviation formula. In our comparison in the result section \ref{sec:results}, we have taken the best case stars N11-036 and N11-109 [$log_{10}(L/L_{\odot})$ = (4.95, 4.48) and $T_{eff}$ = (23,750, 25,750) K, respectively], which have similar observational uncertainties as N11-095. Note that these are closer to the main-sequence rather than in the post-MS supergiant stages, but we anticipate that the uncertainties would be similar for such supergiants.
}
%%%% Table %%%%%%
%\begin{deluxetable*}{ccccccccccc}
%\tablecaption{Surface Chemical Abundances of LMC Supergiants \label{tab:CNO_abund}}
%\tablewidth{0pt}
%\tabletypesize{\scriptsize}
%\tablehead{
%\colhead{Star} & \colhead{Spectral Type} &  \colhead{Teff (K)} & 
%\colhead{$log(L/L_\odot)$} & \colhead{log g} & \colhead{$M/M_\odot$} &
%\colhead{C} & \colhead{N} & \colhead{O} & \colhead{$N/C$} & \colhead{$N/O$} 
%}
%\colnumbers
%\startdata
%N 11-109\tablenotemark{a} & {B0.5 Ib} & {25750}  & {4.48} & {3.2} & {12} & {7.41} & {7.24} & {8.32} & {0.68} & {0.08}  \\
%N 11-110\tablenotemark{a} & {B1 III} & {23100} & {4.37} & {3.25} & {12} & {7.49} & {7.39} & {8.48} & {0.79} & {0.08}\\
%N 11-124\tablenotemark{a} & {B0.5 V} & {28500} & {4.47} & {4.2} & {14} & {7.56} & {7.25} & {8.12} & {0.49} & {0.13}  \\
%RD92 H II\tablenotemark{b} & {--} & {--} & {--} & {--} & {--} & {8.04} & {7.14} & {8.35} & {0.13} & {0.06}   \\
%\enddata
%\tablenotetext{a}{\citet{Hunter:2007aa}}
%\tablenotetext{b}{\citet{Evans:2004aa}}
%\tablecomments{This table is also available in the machine-readable format online. The selection of the data is explained in section \ref{sec:obs_surf_abund}}
%\end{deluxetable*}
%%%%%%%%%%%%%%%%%%%

\section{Results}\label{sec:results}

The post-main sequence BSGs have hydrogen burning shells\deleted{ and have started or are yet to start burning helium in its core} \added{around helium cores that are either inert or have just started to burn helium in their central parts} (see the regions marked in red just before model 1800, $\sim$ 1.16 Myr before CC in Fig. \ref{fig:13M_kipp}). Massive stars in the post-main sequence (post-MS) phase, last for very short duration in comparison with core helium burning. Thus in its first transit from the Terminal Age Main Sequence (TAMS) phase to the RSG branch a BSG spends very little time in this part of the HRD compared to the later similar blue to red transition during a blue loop phase. Hence a typical BSG star in the blue part of the HRD is more likely to be one that has evolved from a previous RSG phase. \deleted{This is revealed by the relative times spent in the evolutionary process as reported in Table \ref{tab:br_ratios}. (The methods for probing a BSG star for its RSG history are discussed in section \ref{sec:blue_loops}.)} The post-MS stars have a \deleted{trace of their} MS history \added{differentiated} through the presence or absence of an Intermediate Convective Zone (ICZ) above the hydrogen burning core which \added{in turn} depends on the treatment of convection, mass loss, etc. \added{in model calculation.} The excitation of g-modes in BSGs has been shown to be related to the ICZ \citep{Dupret:2009aa}. These authors show that radiative damping significantly affects most non-radial modes of the very dense frequency spectrum of BSGs. However, the presence of an ICZ above the hydrogen burning shell allow some of the modes to reflect on it and are thus not damped. Thus it makes asteroseismology amenable as a tool to probe and constrain \deleted{different physical} aspects of the MS and post-MS history and the stellar interior. But significant mass loss from the star or overshooting of the convective zones can prevent the formation of an ICZ during the post-MS phase and can extend the MS phase \citep{Godart:2009aa}. Combined with the fast evolution of the BSG on its first journey to the red, the period variation may be detectable if the time variation of the frequency of the same mode in the dense spectrum of g-modes \citep{Saio:2013aa} can be observed with high enough precision.

\subsection{Blue loop and convective overshoot}

\added{To elaborate the effect of convective overshoot on the existence of the blue loop,} Fig. \ref{fig:13M_kipp} makes a comparison \deleted{of}\added{between} the Kippenhahn diagrams \deleted{of}{for} two cases of a 13 M$_{\odot}$ star with \deleted{two} different values of the overshoot parameter $f = 0.025$ (left) and $f= 0.032$ (right). In both cases the same value of $f_0= 0.005$ is used. \added{The Kippenhahn diagram in Fig. \ref{fig:13M_kipp} shows the succession of different convective burning phases of the star and zones inside the star during initial hydrogen-helium burning phases as well as more advanced phases of nuclear burning. The lifetime of stars from carbon burning onwards becomes much shorter than those at hydrogen or helium burning stages. This is due to higher core temperatures which leads to direct neutrino cooling dominating over energy transport due to radiation. Energy is also transported by convection inside the star.}

As Fig. \ref{fig:13M_HR} shows the overshoot parameter $f = 0.025$ leads to a "blue loop" in the Hertzsprung Russell diagram (HRD) such that the stellar color shifts to the blue (with a correspondingly higher surface temperature $T_{eff}$) after reaching the red giant branch during a hydrogen shell burning phase. The star with overshoot parameter $f = 0.032$ does not shift its color from the red part of HRD, till it undergoes core collapse.

An examination of the internal structures of the two stars (with simulations represented by $f=0.025$ and $f=0.032$) shows that during the core hydrogen burning phases the two stars have nearly identical structure and core development. The overshoot regions, marked in grey, are predictably larger for the latter model compared to the former. \added{The internal structure starts to differ in the two models towards the end of main-sequence phase.}

\subsection{Hydrogen shell burning and Intermediate Convective Zone}

Even the hydrogen shell burning regions (marked in red \added{in Fig. \ref{fig:13M_kipp}}) are similarly behaved in the two evolutionary simulations up to about model number 2400 ($(\tau_{CC}-\tau) \sim$ 1.12 Myr) when the helium core mass grows to about  $3.2 M_{\odot}$. However we note that the Intermediate Convective Zones (ICZ\deleted{) marked in thin near vertical line (in sky blue color) and present around model number 1800}\added{, visible as sky blue color over red in panels (c) and (d) of Fig. \ref{fig:13M_kipp}, present around model 1800 at mass coordinate of 5--7 \msun)} show some differences. The one on the left (\added{panel (c)}, lower f \added{value of 0.025}) has an ICZ reaching deeper towards the centre of the star than \added{the one} on the right (\added{panel (d), higher f value of 0.032}), while the ICZ on the right is more ``frayed" than the one on the left, showing some radiating branches in the mass coordinate - model number space. The redward evolution of the star through the Hertzsprung gap has been related to the extension of the intermediate convection zone in association with the hydrogen burning shell \citep[see e.g.][]{Meynet:2011aa,Stothers:1973aa, Stothers:1991aa}. The density gradients are usually shallower in the convective zone and larger the ICZ is in mass extent more compact and blue is a star on the HRD. When this ICZ decreases in mass extent or disappears altogether, it leads to the expansion of the star on a rapid timescale and leads the star to move to the red part of the HRD. The crossings of the Hertzsprung gap, whether single or triple (depending on whether a blue loop is present) take place on the Kelvin-Helmholtz (thermal) timescale of the envelope.  As noted by \cite{Stothers:1973aa}, before each crossing of the Hertzsprung gap, there is a fluctuation of the luminosity, such as the exhaustion of the fuel at the center and the resultant switch to the release of gravitational energy in the core during the first passage. Thermal instability of the envelope and the reduction of pressure on the core leads to the luminosity fluctuation before the second passage to the red.

\added{Note however that \citet{Eggenberger:2002aa} and \citet{Maeder:2001aa} point out that the inclusion of stellar rotation in modeling leads to mild mixing and can lead to more helium in the region of H-shell burning. This affects the stellar opacity driving it lower and the ICZ becomes less prominent or absent altogether. The absence of ICZ leads to stellar expansion to the supergiant phase. (See also our discussion about \citet{Saio:2013aa} work in section \ref{subsec:blue_loop})}

%%%% Figure %%%%%%
\begin{figure*}[htb!]
\gridline{\fig{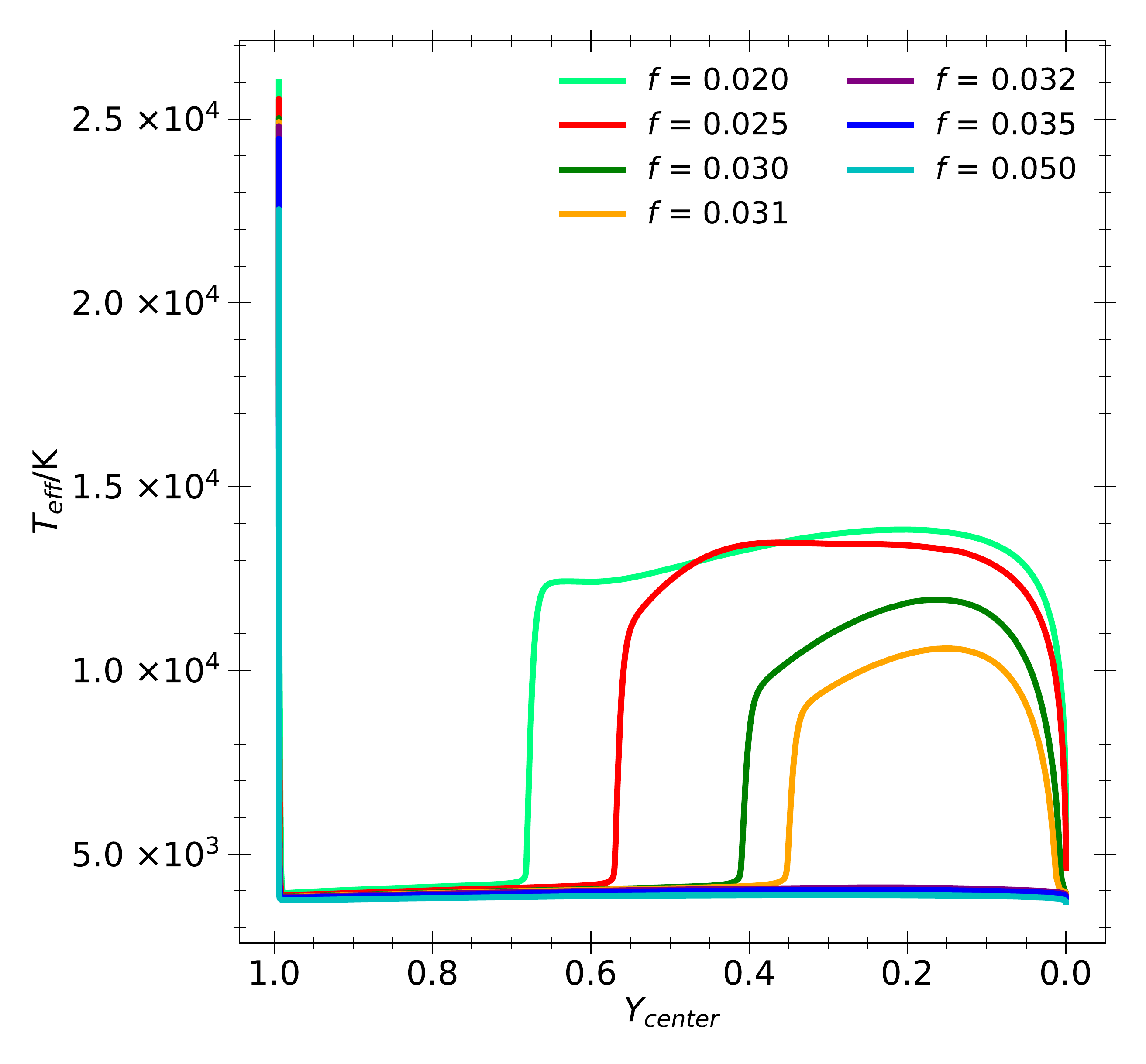}{0.49\textwidth}{(a)}
				\fig{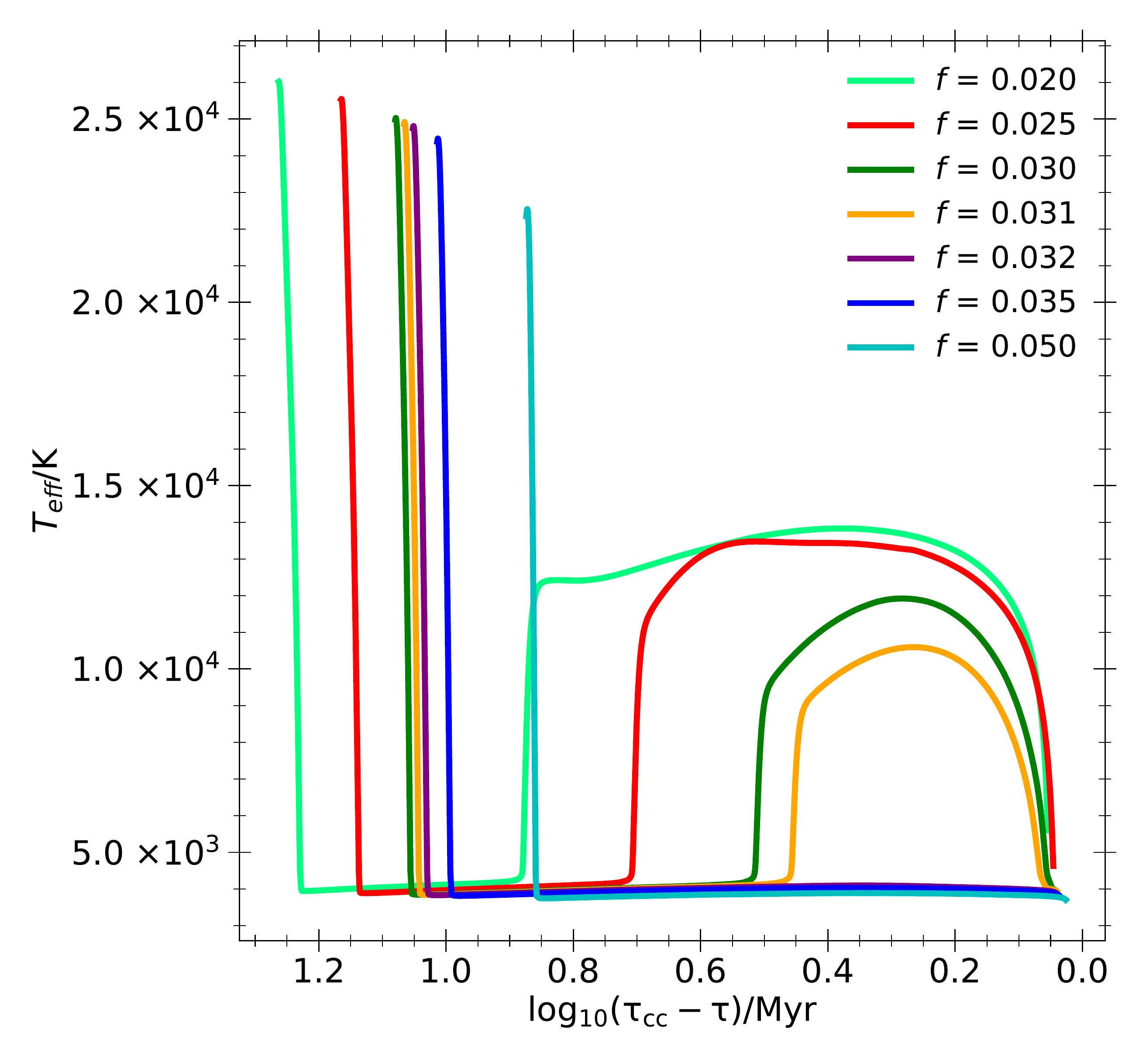}{0.49\textwidth}{(b)}}
\caption{\Teff \ as a function of central helium mass fraction ($\rm Y_c$) \added{is shown in panel (a), while \Teff \ evolution with time before collapse (in Myr) is shown in panel (b)} for M$_{ZAMS}$= 13 \msun \  and Z = 0.006 models with different values of overshoot parameter $f$ with $f_0$ held constant at 0.005. \added{The plots display the evolution post-MS when $\rm Y_c$ is unity through the helium burning until central helium is completely exhausted. The models that exhibit a blue loop ($f \le$ 0.031) show a transition in \Teff \ from red towards blue, while for the other models ($f >$ 0.031) \Teff \ remains in red after the first blue to red transition post-MS. The time spent in helium burning phase, and blue loop (when it exists, along with maximum temperature excursion in blue) differs depending on the value of $f$.}
\label{fig:13M_Teff_Yc}}
\end{figure*}
%%%% %%%%%%%%%%

\subsection{Contrast between the outer convective zones for different overshoot parameters}

By the time helium is ignited in the core (\added{about} model 2000 \added{in panels (c) and (d) or} $(\tau_{CC}-\tau) \sim$ 1 Myr \added{in panels (a) and (b) of Fig. \ref{fig:13M_kipp}}), a surface convective zone begins to form and extends rapidly deeper into the star in both model simulations \added{with $f=0.025$ and $f=0.032$ presented} in Fig. \ref{fig:13M_kipp}. Thereafter, \deleted{the helium burning convective core growth, the hydrogen shell-burning evolution and, most importantly, }the development of the outer convective region in the hydrogen rich envelope (so-called "non-burn" envelope) begin to differ markedly\deleted{ for the two simulations marked by $f=0.025$ and $f=0.032$}. \added{Noticeable differences are also seen in the helium burning convective core growth and the hydrogen shell burning evolution.}  For the \deleted{former }case \added{of $f$ = 0.025}, the outermost convective zone recedes rapidly to the surface by about model 2800 [$(\tau_{CC}-\tau) \sim$ \e{7}{5} yr], leaving only a radiative mantle and envelope beyond the hydrogen burning shell. The radiative envelope persists till about model 3300 [$(\tau_{CC}-\tau) \sim$ \e{4.98}{4} yr] after which a small outer convective zone forms during the late core helium burning phase. This small convective envelope then again rapidly grows deep down to make the star fully convective beyond the boundary of the helium core. 

In contrast, for the model with higher overshoot factor $f=0.032$, which does not undergo a blue loop in the HRD, while there is a narrowing of the outer convective envelope starting at model number 2400 [($\tau_{CC}-\tau) \sim$ 1.12 Myr] for another 400 models ($\sim$ \e{4.13}{5} yr), leading to a more extended radiative mantle in between, the star never has a fully radiative mantle plus envelope extending all the way to the surface. That is the outer regions of the star in this simulation always retains a outer convective structure. The blue loop therefore appears to be connected with a fully radiative envelope in the model simulation with $f=0.025$. The higher overshooting model also has somewhat more noticeable "undershoot" of the non-burning convective envelope during models 2800 to 3800 [$(\tau_{CC}-\tau) \sim$ \e{7}{5} yr to \e{4}{4} yr]. In the later phases of more advanced nuclear burning, the higher overshoot simulation $f=0.032$ has a slightly higher helium and carbon core masses (4.71 \msun \ \& 2.38 \msun, respectively, compared to 4.4 \msun \ \& 2.19 \msun \ in $f=0.025$ simulation). The mass coordinates of the helium burning shells also extend over a larger range in these phases in the higher overshoot model but only marginally so.

Hydrogen shell burning during the core helium burning phase extends to a maximum mass of 5.4 $M_{\odot}$ for $f=0.025$ instead of the maximum mass of 5.1 $M_{\odot}$ for  $f=0.032$. There is a smoother growth of the helium core boundary in the case of $f=0.032$ than compared to several spurts of the helium core growth in the case of $f=0.025$. In both cases semi-convection is in operation near the convective He-core edge. Another point to note from Fig. \ref{fig:13M_kipp} is that for the higher overshoot simulation the star ends up with a lower final pre-supernova star mass than the corresponding mass for the lower overshoot simulation. As marked in this Figure by a black dashed line\deleted{at the top, (12.51 \msun \ in Fig. \ref{fig:13M_kipp}(a) instead of 12.06 \msun \ in Fig. \ref{fig:13M_kipp}(b))}\added{, these masses are 12.51 \msun \ (Fig. \ref{fig:13M_kipp} (a)) and 12.06 \msun \ (Fig.  \ref{fig:13M_kipp} (b)) at CC}. This is because the higher overshoot simulation has the star stay always in the red part of the HRD once it arrives there \added{post-MS, until CC}. For a RSG star, the mass loss-rate is higher than for blue stars of approximately the same luminosity. Hence a star (with a higher convective overshoot) staying in the RSG branch for most of its advanced nuclear burning phases\deleted{ as opposed to the stars that make the red $\rightarrow$ blue $\rightarrow$ red transition} will end up losing more mass and therefore be somewhat less massive at the pre-SN stage \added{than the star making the red $\rightarrow$ blue $\rightarrow$ red transition (blue-loop)}.

%%%% Figure %%%%%%
\begin{figure}[htb!]
\includegraphics[width=0.48\textwidth]{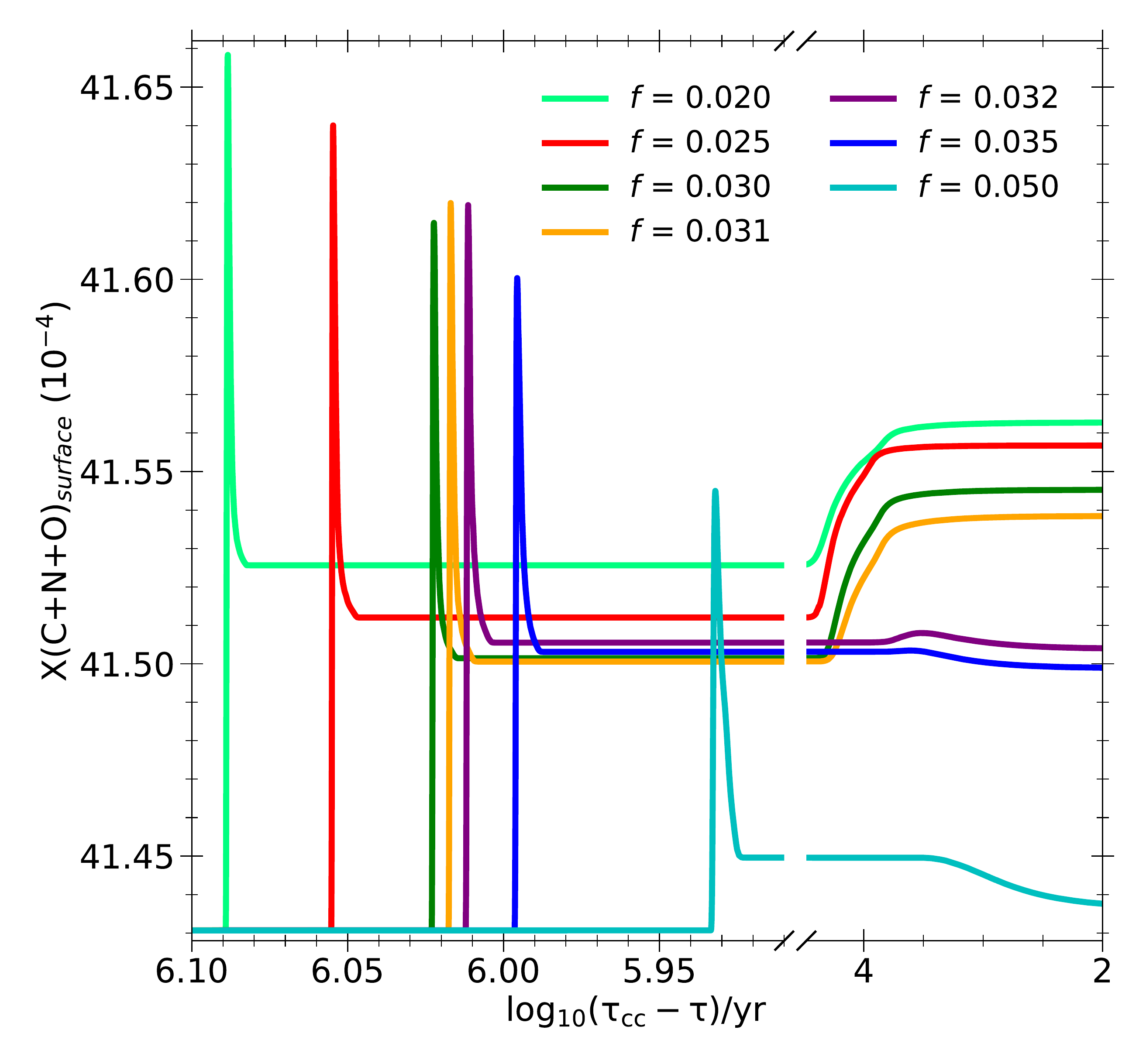}
\caption{The evolution of total surface CNO mass fraction predicted by MESA with time before CC for M$_{ZAMS}$= 13 \msun \  and Z = 0.006 models for different values of overshoot parameter $f$ with $f_0$ = 0.005. \added{Typical observational error bars are discussed in section \ref{sec:obs_surf_abund}.}
\label{fig:13M_CNO}}
\end{figure}
%%%%%%%%%%%%%%

%%%%% Online Table %%%%%%
\begin{deluxetable*}{hlcccccc}[htb!]
\tablecaption{%\textbf{\textcolor{magenta}{
Characteristics of%}} 
blue loops at various stages for 13 \msun \ models \label{tab:time_in_loop}}
%\tablewidth{0pt}
%\tabletypesize{\scriptsize}
\tablehead{
\nocolhead{overshoot} & \colhead{Physical} & \colhead{$log_{10} (T_{eff})$ = 4.2} &
\colhead{$\rm 1^{st}$ min.} & \colhead{$\rm 2^{nd}$ min.} &
\colhead{blue-most point} & \colhead{$\rm 3^{rd}$ min.} &\colhead{Core} \\[-1.5ex]
\nocolhead{parameter} & \colhead{Quantity} & \colhead{post-MS} & 
\colhead{in $\rm L_{surf}$} & \colhead{in $\rm L_{surf}$} & 
\colhead{in blue loop} & \colhead{in $\rm L_{surf}$} & \colhead{Collapse} \\[-1.5ex]
\nocolhead{f} & \colhead{} & \colhead{$\rm t_{0}$} &
\colhead{$\rm t_{1}$} & \colhead{$\rm t_{2}$} & 
\colhead{$\rm t_{3}$} & \colhead{$\rm t_{4}$} & 
\colhead{$\rm t_{5}$} 
}
%\colnumbers
\startdata
\multicolumn{7}{c}{$f = $ 0.020} \\
\hline
\multirow{4}{1cm}{0.020} & model number & 1886 & 2183 & 2532 & 2909 & 3710 & 138753 \\
& $\rm t - t_{CC} [10^3 yrs]$ & 1241.30 & 1229.01 & 936.74 & 376.22 & 50.53 & 0.00 \\% \e{1.24}{6} & \e{1.23}{6} & \e{9.37}{5} & \e{3.76}{5} & \e{5.05}{4} & 0 \\
& $\rm L_{surf}$ [$10^3$\Lsun] & 38.57 & 21.91 & 35.53 & 49.49 & 33.09 & 68.79 \\% \e{3.86}{4} & \e{2.19}{4} & \e{3.55}{4} & \e{4.95}{4} & \e{3.31}{4} & \e{6.88}{4} \\
& \Teff [$10^3$K] & 15.89 & 4.41 & 4.16 & 13.83 & 4.28 & 3.82 \\%\e{1.59}{4} & \e{4.41}{3} & \e{4.16}{3} & \e{1.38}{4} & \e{4.28}{3} & \e{3.82}{3} \\
\hline
\multicolumn{7}{c}{$f = $ 0.025} \\
\hline
\multirow{4}{1cm}{0.025} & model number & 1892 & 2190 & 2588 & 2921 & 3706 & 113682 \\
& $\rm t - t_{CC} [10^3 yrs]$ & 1147.25 & 1136.83 & 766.69 & 513.56 & 43.74 & 0.00 \\% \e{1.15}{6} & \e{1.14}{6} & \e{7.67}{5} & \e{5.14}{5} & \e{4.37}{4} & 0 \\
& $\rm L_{surf}$ [$10^3$\Lsun] & 40.35 & 23.62 & 40.33 & 51.96 & 33.76 & 77.52 \\%\e{4.04}{4} & \e{3.83}{4} & \e{4.03}{4} & \e{5.20}{4} & \e{3.38}{4} & \e{7.75}{4} \\
& \Teff [$10^3$K] & 15.86 & 4.39 & 4.12 & 13.48 & 4.28 & 3.78 \\%\e{1.59}{4} & \e{1.00}{4} & \e{4.12}{3} & \e{1.35}{4} & \e{4.28}{3} & \e{3.78}{3} \\
\hline
\multicolumn{7}{c}{$f = $ 0.030} \\
\hline
\multirow{4}{1cm}{0.030} & model number & 1898 & 2193 & 2669 & 2982 & 3395 & 100650 \\
& $\rm t - t_{CC} [10^3 yrs]$ & 1064.73 & 1055.60 & 608.85 & 283.82 & 51.29 & 0.00 \\%\e{1.06}{6} & \e{1.06}{6} & \e{6.09}{5} & \e{2.84}{5} & \e{5.13}{4} & 0 \\
& $\rm L_{surf}$ [$10^3$\Lsun] & 42.42 & 26.15 & 45.55 & 55.94 & 42.75 & 84.83 \\%\e{4.24}{4} & \e{2.61}{4} & \e{4.55}{4} & \e{5.59}{4} & \e{4.27}{4} & \e{8.48}{4} \\
& \Teff [$10^3$K] & 15.83 & 4.37 & 4.09 & 11.92 & 4.17 & 3.75 \\%\e{1.58}{4} & \e{4.37}{3} & \e{4.09}{3} & \e{1.19}{4} & \e{4.17}{3} & \e{3.75}{3} \\
\hline
\multicolumn{7}{c}{$f = $ 0.031} \\
\hline
\multirow{4}{1cm}{0.031} & model number & 1901 & 2195 & 2698 & 2994 & 3363 & 92009 \\
& $\rm t - t_{CC} [10^3 yrs]$ & 1051.54 & 1042.58 & 568.41 & 263.63 & 60.78 & 0.00 \\% \e{1.05}{6} & \e{1.04}{6} & \e{5.68}{5} & \e{2.64}{5} & \e{6.08}{4} & 0 \\
& $\rm L_{surf}$ [$10^3$\Lsun] & 42.96 & 26.77 & 46.77 & 56.44 & 44.81 & 86.62 \\%\e{4.30}{4} & \e{2.68}{4} & \e{4.68}{4} & \e{5.64}{4} & \e{4.48}{4} & \e{8.66}{4} \\
& \Teff [$10^3$K] & 15.89 & 4.36 & 4.08 & 10.60 & 4.16 & 3.74 \\%\e{1.59}{4} & \e{4.36}{3} & \e{4.08}{3} & \e{1.06}{4} & \e{4.16}{3} & \e{3.74}{3} \\
\hline
\multicolumn{7}{c}{$f = $ 0.032} \\
\hline
\multirow{4}{1cm}{0.032} & model number & 1900 & 2193 & \multicolumn{3}{c}{\multirow{4}{*}{No blue loop}} & 90275 \\ 
& $\rm t - t_{CC} [10^3 yrs]$ & 1038.05 & 1029.45 %\e{1.04}{6} & \e{1.03}{6} 
& & & & 0.00 \\
& $\rm L_{surf}$ [$10^3$\Lsun] & 43.13 & 27.40  %\e{4.31}{4} & \e{2.74}{4} 
& & & &  87.25 \\ %& \e{8.73}{4} \\
& \Teff [$10^3$K] & 15.83 & 4.36 %\e{1.58}{4} & \e{4.36}{3} 
& & & &  3.73 \\%& \e{3.73}{3} \\
\hline
%\multicolumn{7}{c}{$f = $ 0.033} \\
%\hline
%\multirow{4}{1cm}{0.033} & model number & 1905 & 2198 & \nodata & \nodata & \nodata & 94132 \\
%& $\rm t - t_{CC} [10^3 yrs]$ & \e{1.03}{6} & \e{1.02}{6} & \nodata & \nodata & \nodata & 0 \\
%& $\rm L_{surf}$ [$10^3$\Lsun] & \e{4.35}{4} & \e{2.79}{4} & \nodata & \nodata & \nodata & \e{8.97}{4} \\
%& \Teff [$10^3$K] & \e{1.59}{4} & \e{4.35}{3} & \nodata & \nodata & \nodata & \e{3.72}{3} \\
%\hline
\multicolumn{7}{c}{$f = $ 0.035} \\
\hline
\multirow{4}{1cm}{0.035} & model number & 1908 & 2198 & \multicolumn{3}{c}{\multirow{4}{*}{No blue loop}} & 82520\\
& $\rm t - t_{CC} [10^3 yrs]$ & 1000.96 & 992.96 % \e{1.00}{6} & \e{9.93}{5} 
& & & & 0.00 \\
& $\rm L_{surf}$ [$10^3$\Lsun] & 44.35 & 29.04 %\e{4.43}{4} & \e{2.90}{4} 
& & & & 92.85 \\%& \e{9.28}{4}\\
& \Teff [$10^3$K] & 15.88 & 4.35 %\e{1.59}{4} & \e{4.35}{3} 
& & & & 3.71 \\%& \e{3.71}{3}\\
\hline
\multicolumn{7}{c}{$f = $ 0.050} \\
\hline
\multirow{4}{1cm}{0.050} & model number & 1948 & 2227 & \multicolumn{3}{c}{\multirow{4}{*}{No blue loop}} & 74868 \\
& $\rm t - t_{CC} [10^3 yrs]$ & 864.72 & 859.06 %\e{8.65}{5} & \e{8.59}{5} 
& & & & 0.00 \\
& $\rm L_{surf}$ [$10^3$\Lsun] & 51.64 & 38.26 %\e{5.16}{4} & \e{3.83}{4} 
& & & & 121.59 \\%& \e{1.21}{5}\\
& \Teff [$10^3$K] & 15.66 & 4.30 %\e{1.57}{4} & \e{4.30}{3} 
& & & & 3.63 \\%& \e{3.63}{3}\\
\enddata
\tablecomments{Physical quantities at specific points during the evolution of the star for the  M$_{ZAMS} $ = 13 \msun , Z = 0.006 models are listed here. %\textbf{\textcolor{magenta}{
Model number is the number of time steps taken in the MESA simulation since the ZAMS stage. %}} 
$\rm t_{0}$ is the point where the star transitions red-ward post-TAMS phase and crosses \lgTeff \ = 4.2 for the first time. $\rm t_{1}$, $\rm t_{2}$, and $\rm t_{4}$ are \deleted{points}%\textbf{\textcolor{magenta}{
three local minima in the HRD (refer to Fig. \ref{fig:13M_HR}) %}}
where the luminosity reaches a minimum value before it rises again\deleted{ in the HRD (ref. figure \ref{fig:13M_HR})} for each consecutive blue-ward and red-ward transition %\textbf{\textcolor{magenta}{
in the models that exhibit a blue loop ($f \le$ 0.031). In other models without a blue loop, there is only one luminosity minimum at $\rm t_{1}$. For the models that exhibit a blue loop, %}} 
$\rm t_{3}$ is the point in blue loop where the \Teff \ becomes maximum, before the  %\textbf{\textcolor{magenta}{
evolutionary %}}
track turns red-ward again. The %\textbf{\textcolor{magenta}{
point %}} 
$\rm t_{5}$ is %\textbf{\textcolor{magenta}{
time of core-collapse, %}} 
$\rm t_{CC}$.}
\end{deluxetable*}
%%%%%%%%%%%%%%%%%%%%

\subsection{Maximum temperature in the Blue Loop}

\added{
In Fig. \ref{fig:13M_Teff_Yc} we have shown the evolution of  \Teff \ as a function of central helium abundance in panel (a) (which starts out at unity, since this is post-MS evolution, and stops when helium is completely exhausted in center of the star), and \Teff \  as a function of time before CC in panel (b). The 13 \msun \ star makes a rapid transition across the  Hertzsprung Russell gap (vertical blue to red transition on the left-most edge in panels (a) and (b)) for all overshoot factors. A few of these stars (simulations with overshoot parameter $f \le$ 0.031) however make a single blue loop leading to higher \Teff \ at intermediate times before switching back to red again. These temperature excursions as well as the starting point and duration of these blue loops depend upon the overshoot factors $f$. As noted earlier, beyond $f=0.31$, there is no further blueward excursion of the \Teff \ (blue-loop) and these stars as well as the stars that undergo the blue loops explode finally as RSGs.
}

The maximum temperature that the star achieves in its blueward loop is also affected by the convective overshoot parameter.  %\added{These are given in Table \ref{tab:time_in_loop} column 6, last of the 4 rows for each overshoot factor f. Additionally, the time spent in the blue loop ($\tau_{B_2}$) is also given in column 6 of Table \ref{tab:br_ratios}, for each f, which are later used in equation \ref{eqn:BtoR}.} 
Typically the blueward extension goes to a maximum of  \lgTeff \ $\approx$ 4.14  (\Teff \ $\approx$ 13,800 K \added{with $L_{surf} \approx$ 49,500 \Lsun}) for the parameter $f = 0.020$ and $f_0 = 0.005$  for the 79 isotopes network. \added{The maximum temperature in the blue loop decreases, while the surface luminosity increases as the value of overshoot parameter $f$ increases. Luminosity and Temperature at the maximum excursion in the blue loop are listed in column `t$_3$' of Table \ref{tab:time_in_loop}.} \added{A simulation (not presented in this paper) run} using input parameters $f$= 0.004 and $f_0$ = 0.001, and a 22-isotopes network \citep[choices made by][]{Farmer:2016aa}, the maximum \lgTeff \ $\approx$ 4.25  (\Teff \ $\approx$ 17700 K \added{with $L_{surf} \approx$ 42,200 \Lsun}) in the blue loop. This range of temperatures are partially covered by the LMC supergiants as seen in Fig. \ref{fig:13M_HR}.

%%%% Figure %%%%%%
%\begin{figure}[htb!]
%\plotone{Fig2.pdf}
%\caption{The ratio of total nitrogen to total carbon abundance (including all isotopes) is plotted against the ratio of total nitrogen to total oxygen abundance at the surface throughout the history of the stars evolution for M$_{ZAMS}$= 13 \msun ,  Z = 0.006 models for different values of overshoot parameter $f$ with $f_0$ held constant at 0.005. The data for LMC supergiants is also plotted in the figure (see section \ref{sec:obs_surf_abund} for details).
%\label{fig:13M_NC_NO}}
%\end{figure}

\begin{figure*}[htb!]
\gridline{\fig{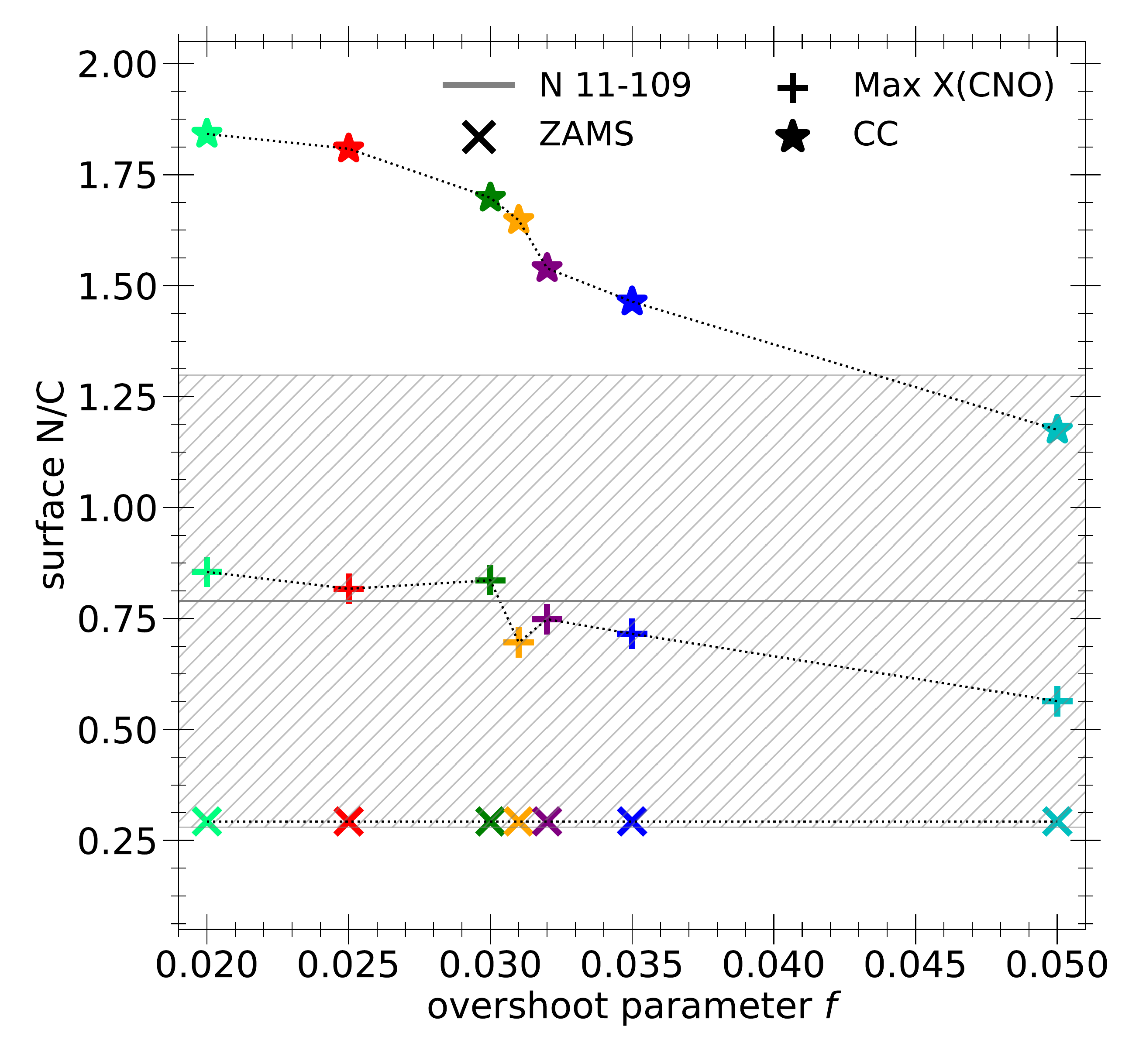}{0.49\textwidth}{(a)}
				\fig{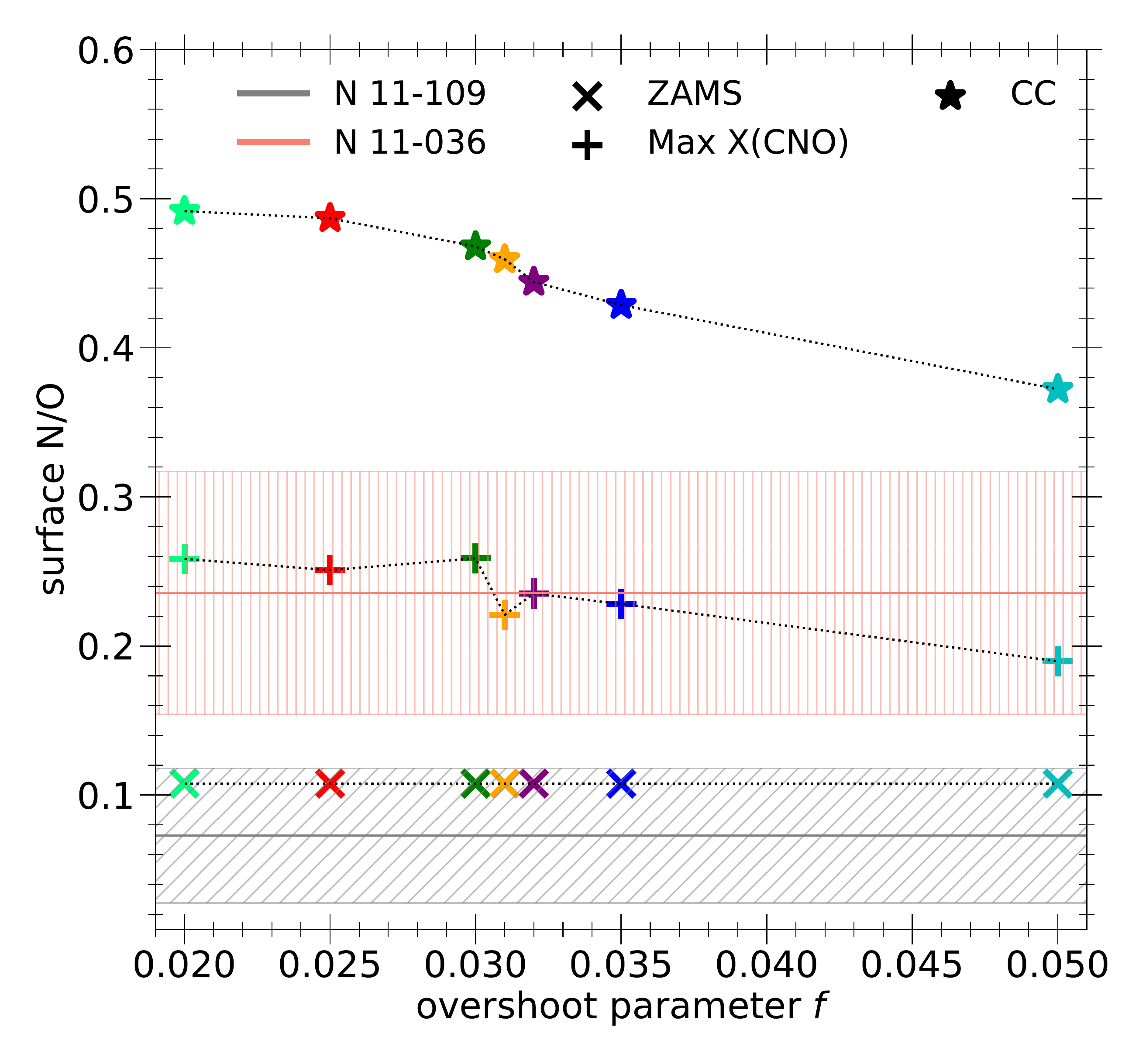}{0.49\textwidth}{(b)}}
\caption{\added{The ratio of total surface nitrogen to total surface carbon mass fractions (including all isotopes) in panel (a). The ratio of total surface nitrogen to total surface oxygen mass fractions (including all isotopes) in panel (b). All these ratios are shown at three different stages in the evolution for M$_{ZAMS}$= 13 \msun ,  Z = 0.006 models for different values of overshoot parameter $f$ with $f_0$ = 0.005. In both panels, the points are joined a dotted line for visual clarity. The shaded regions show selected observations for LMC supergiants from \citet{Hunter:2007aa} with uncertainties (see section \ref{sec:obs_surf_abund}).}
\label{fig:13M_NC_NO_vs_f}}
\end{figure*}
%%%%%%%%%%%%%%

\subsection{Surface chemical enrichment}

\deleted{The external convection zone (when present) dredges out the CNO products from the hydrogen shell burning layers to the surface, so that these elements may have a different abundance and their \added{mutual} ratios than in the primordial gas from which the star formed. Similarly, if the external convection zone thoroughly mixes elements from the hydrogen burning shell at an earlier stage, the effect of later enhancement from a subsequent dredge-up may be masked. The presence of external convection zone in the star (or the lack thereof, even if for a limited duration and in intermediate models) may therefore affect the surface abundance of these elements. The overshoot factor, as we have seen above, is a controlling factor for the extent of convection zone. }

We show in Fig. \ref{fig:13M_CNO} how the abundance of different CNO nuclei at the surface evolves with time before core-collapse as a function of the overshoot factor. \deleted{We note that while a trend in simulation is discernible it may be hard to distinguish abundance differences observationally. }The surface CNO abundance undergoes enrichment soon after helium ignition in the core (see Fig. \ref{fig:13M_CNO}). At this time onward the external convection zone rapidly grows inward (see \deleted{the left panels of }Fig. \ref{fig:13M_kipp}). The external convection zone dredges out the CNO nuclei generated in the Hydrogen burning shell above the helium core. For $f = 0.025$ (red curve in Fig. \ref{fig:13M_CNO}), the initial rise in the CNO curve, at $\sim$1 Myr before CC, lasts for about 20,000 yr (3000 yr to the maximum of X(CNO) followed by 17000 yr fall to the intermediate plateau). In the Kippenhahn diagram \added{(Fig. \ref{fig:13M_kipp}, panel(c))} this corresponds to model range 2200 to 2400 with the largest penetration of the external convection zone occurring at about model number 2380 (after which the external convection zone recedes to the surface). The elapsed time from the Kippenhahn plot (\deleted{left panel}\added{panel(a)}) is 19835 yr. The intermediate plateau in Fig. \ref{fig:13M_CNO} lasts for about 1.1 million years, \deleted{between model numbers 2400 and 3800, }which is during the time when the external convection boundary moves out to the surface. This convection remains shut for sometime and then again penetrates deeper to reach near the boundary of the hydrogen burning shell. After the intermediate plateau, the CNO abundance rises again to a newer plateau for all cases that undergo a blue loop.

The highest enrichment factor for CNO is higher for $f=0.025$ than for $f= 0.033 \; \rm or \; 0.050$. \deleted{This is due to a rapid dredge-up of the CNO nuclei and less efficient dilution mixing into the envelope of the original composition of the star for a shorter duration for the lower overshoot factor $f=0.025$.They all}\added{The CNO abundances, all} come down to an intermediate plateau which too depends on the overshoot factor $f$ and the plateau values are \deleted{ordered by the peak values} reached around $(\tau_{CC} - \tau) \sim$ \powten{6} yr. Thereafter, about $3 \times 10^4$ yr before core collapse, around carbon core growth and shell helium burning stages a further smaller enrichment episode occurs which persists till the core collapse stage. Both these enrichment phases are assisted by the external convection zone that spans to the surface of the star. Only the stars that show a blue loop in the HRD have a substantial second phase enrichment\footnote{The second rise in red curve ($f=0.025$) lasts for about 27,000 years between model numbers 3200 and 3800.This is during the phase when the external convection zone penetrates deeply again towards the now nearly extinguished hydrogen shell at the top of the helium core.}. The others that do not undergo a full blue-loop have either marginal second phase enrichment or an actual fall (for $f=0.050$) from the intermediate plateau to a new plateau that lasts till core collapse. For example, in the model with $f = 0.032$\deleted{ (part (b) of Fig. \ref{fig:13M_kipp})}, the magenta curve in \deleted{X(CNO),}\added{Fig. \ref{fig:13M_CNO}} has a second rise not as dramatic \added{as a lower $f$ value,} because a large external convection was already present \added{(see panels(b) \& (d) of Fig. \ref{fig:13M_kipp})}, although its extent was limited for some time (between models 2400 and 3800). \added{While a trend in simulation as seen in Fig. \ref{fig:13M_CNO} is discernible, the variation in X(CNO) from ZAMS stage to the maximum value for $f$ = 0.020 is less than 1\%. Thus, these variations are at present observationally indistinguishable (see section \ref{sec:obs_surf_abund})}.

\deleted{The enrichment of surface Nitrogen also has to be accompanied by correlated changes of Carbon and Oxygen as they are produced in the CNO cycle. The plots of the ratios of (N/C) vs (N/O) are sensitive indicators of the internal mixing of massive stars \cite{Maeder:2014aa}. In Fig. \ref{fig:13M_NC_NO}, we show the (N/C) vs (N/O) ratios of both 13 \msun \ (for different values of  overshoot factors) along with a few observational points%\footnote{\citet{Maeder:2014aa} confirm some CNO products mixing for supergiant stars in the LMC but they argue that the large scatter in the observational points is due to the insufficient accuracy of the data for testing the various models of stars. They reanalyzed a fraction of the early B-type stars in the VLT-FLAMES data set of \citet{Hunter:2008aa,Hunter:2009aa} using their own spectral line formation computations, and gave arguments for why part of the observational data should not be used for this comparison. Their verification of ionization equilibria and abundance determination (see their Table 1) showed that rather few of their results made a good match of abundances of He \textsc{i}, He \textsc{ii}, C, N, O, Mg \textsc{i} and Si \textsc{iii}, \textsc{iv}. Among their stars listed under LMC (9 in all) several are showing double lined spectra (spectroscopic binaries) or are Be stars. Among the normal stars, only one, N11-095 have both He \textsc{i}, C \textsc{ii}, N \textsc{ii}, O \textsc{ii} determined. Unfortunately this star is too blue ($T_{eff}= 26,800 K$) to be of consequence for our comparison in the $L - T_{eff}$ space.} 
(see subsection \ref{sec:obs_surf_abund} for details). These curves have an upward curvature.}

\added{The enrichment in CNO abundance is dominated by the enrichment in nitrogen (about a factor of 4 between the ZAMS and the CC stages). In Fig. \ref{fig:13M_NC_NO_vs_f}, we show the ratios of surface nitrogen to carbon mass fractions (surface (N/C), panel (a)) and surface nitrogen to oxygen mass fractions (surface (N/O), panel (b)) as reported by MESA. Both these ratios are plotted for different $f$ values at three distinct stages through the evolution of the star. The bottom-most values are the ZAMS values, which are same for all $f$ values, as expected. The values in the middle are ratios at the evolutionary stage when the surface X(CNO) is at maximum for the given $f$ values (around \powten{6} years before CC in Fig. \ref{fig:13M_CNO}). The top-most values are the ratios at CC.  We have also shown abundance ratios in select stars as the best case examples (least uncertainties in the observations) in the figure. As we can see that even for this best case, observed (N/C) ratio cannot be used to distinguish between different evolutionary stages or $f$ values. This is because the uncertainties in determining the C abundance are large. However, we find that the observed ratio (N/O) can be used to marginally differentiate between different evolutionary stages, and perhaps between the extreme $f$ values at the CC stage.  With higher precision observational determination of these ratios in future with larger telescopes and better spectrographs and modeling of spectra, it may be eventually possible to discriminate between different overshoot predictions.}

%The plots of the ratios of (N/C) vs (N/O) are sensitive indicators of the internal mixing of massive stars \cite{Maeder:2014aa}. In Fig. \ref{fig:13M_NC_NO}, we show the (N/C) vs (N/O) ratios of both 13 \msun \ (for different values of  overshoot factors) along with a few observational points\footnote{\citet{Maeder:2014aa} confirm some CNO products mixing for supergiant stars in the LMC but they argue that the large scatter in the observational points is due to the insufficient accuracy of the data for testing the various models of stars. They reanalyzed a fraction of the early B-type stars in the VLT-FLAMES data set of \citet{Hunter:2008aa,Hunter:2009aa} using their own spectral line formation computations, and gave arguments for why part of the observational data should not be used for this comparison. Their verification of ionization equilibria and abundance determination (see their Table 1) showed that rather few of their results made a good match of abundances of He \textsc{i}, He \textsc{ii}, C, N, O, Mg \textsc{i} and Si \textsc{iii}, \textsc{iv}. Among their stars listed under LMC (9 in all) several are showing double lined spectra (spectroscopic binaries) or are Be stars. Among the normal stars, only one, N11-095 have both He \textsc{i}, C \textsc{ii}, N \textsc{ii}, O \textsc{ii} determined. Unfortunately this star is too blue ($T_{eff}= 26,800 K$) to be of consequence for our comparison in the $L - T_{eff}$ space.} (see subsection \ref{sec:obs_surf_abund} for details). These curves have an upward curvature.

%%%% Figure %%%%%%
\begin{figure*}[htb!]
\plottwo{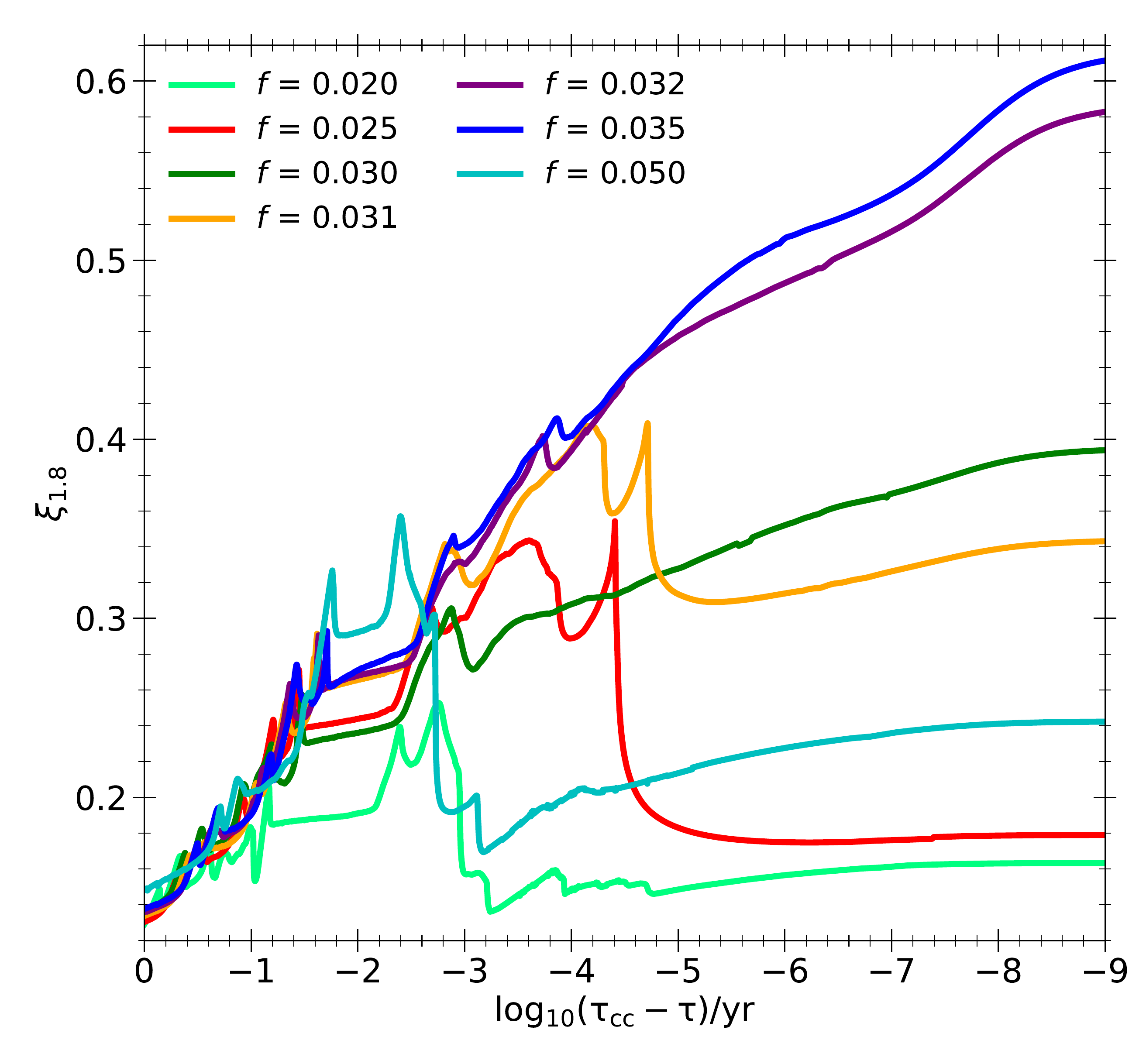}{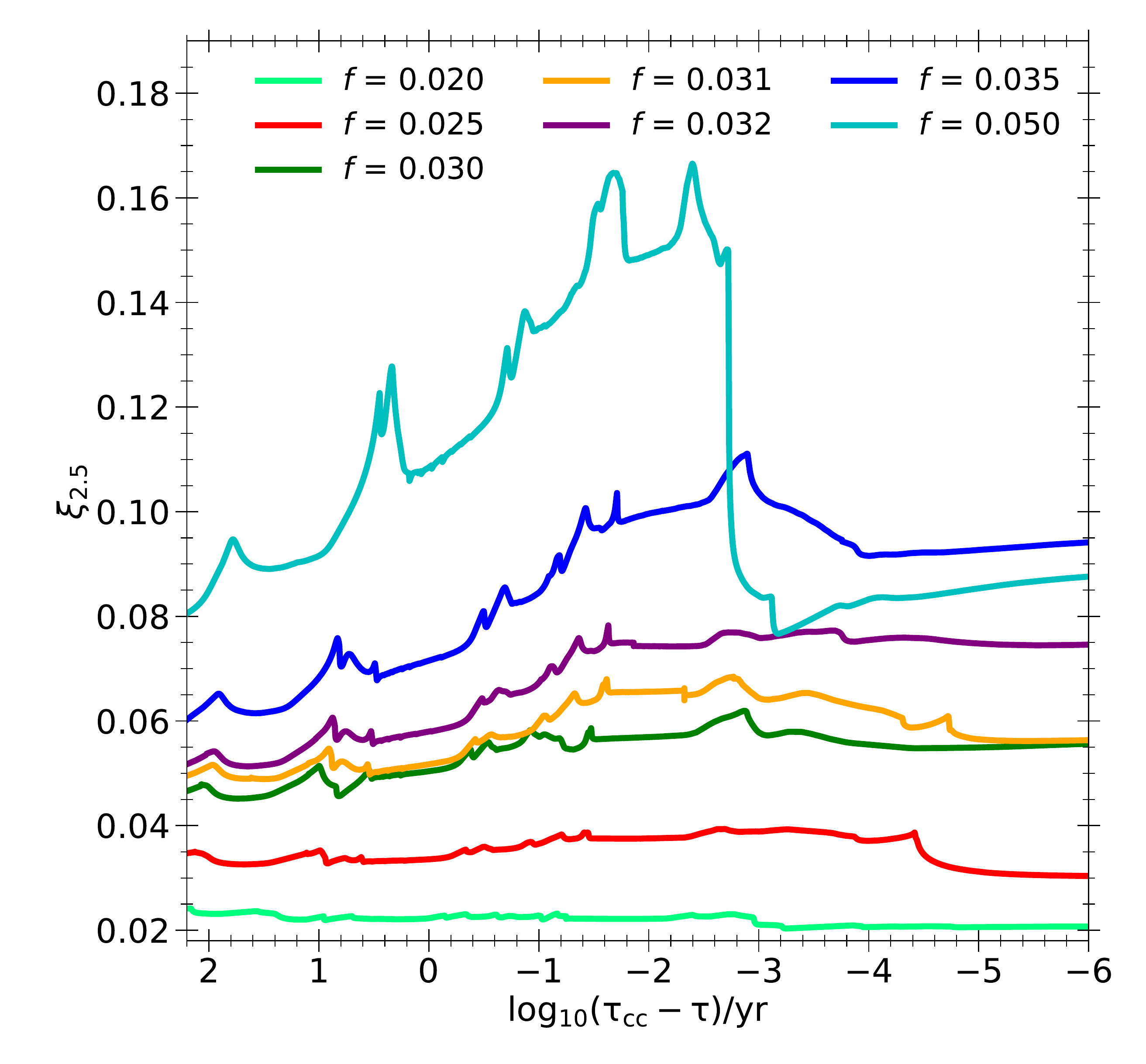}
\caption{Comparison of the evolution of the compactness parameter $\xi_M$ value evaluated at M=1.8\msun \ and M=2.5\msun \ for the M$_{ZAMS}$= 13 \msun \  and Z = 0.006 models for different values of overshoot parameter $f$ with $f_0$ held constant at 0.005.  The value of the compactness parameter at CC is a measure of the ``explodability" of the progenitor star.
\label{fig:13M_compactness}}
\end{figure*}
%%%%%%%%%%%%%%

%\subsection{\sout{\textbf{\textcolor{red}{Blue to red transition and B/R supergiants ratio}}}}

\deleted{We note that in the VLT-FLAMES survey, while most of the Galactic stars were members of central clusters, the LMC and SMC stars were predominantly fields stars in the Magellanic Clouds. Since the field stars are not co-eval like those in the clusters, we can assume a stellar population in steady state with a constant star formation rate. We can thus evaluate the BSG to RSG ratio in terms of the relative duration of the two supergiant phases. For constant star formation rate, the blue to RSG ratio B/R can be written as \citep{Ekstrom:2013aa}:
%
%\begin{equation} \label{eqn:BtoR}
%\frac{B}{R} = \frac{\int \tau_B \phi(m) dm}{ \int \tau_R \phi(m) dm}
%\end{equation}
%
where $\phi(m)$ is the stellar mass distribution function and the integral is carried over the mass range of supergiant stars, namely from 10 \msun \ to about 25 \msun . %\added{The lifetimes spent by the star in the BSG or the RSG branch are given in Table \ref{tab:br_ratios}, while the lifetime in blue-loops can be obtained by taking difference in timescales given in columns t$_4$ and t$_2$ (columns 5 and 7) of Table \ref{tab:time_in_loop} }. 
That the B/R ratio depends on the overshooting parameters can be seen from the entries in Table \ref{tab:br_ratios} where the ratio is tabulated for single stellar mass (e.g. 13 \msun). For a single stellar mass the integration over dm in the B/R ratio expression above reduces to the ratio of the lifetimes.
%On the basis of these lifetimes tabulated in Table \ref{tab:br_ratios}, the blue vs RSG ratio B/R can be calculated as:
%$$B/R = {\int \tau_B \phi(m) dm \over \int \tau_R \phi(m) dm}$$
%
The B/R ratio decreases by almost a factor of 4 as the overshooting factor $f$ increases from $f= 0.020$  to $f=0.031$ for the cases in 13\msun \ star which undergo blue loops. 
Although our chosen metallicity $Z = 0.006$ is intermediate between the corresponding metallicities of LMC and SMC (0.007 and 0.002 respectively) we note that the B/R ratio = 0.6 quoted by \citet[][see their Table 3 entry for Spectroscopic ratio for SMC]{Eggenberger:2002aa} is consistent with the range of B/R ratios that we list in the last column of Table \ref{tab:br_ratios} (for the range of overshoot factors that give rise to Blue Loops).
}

\deleted{
In contrast, the Yellow supergiant  vs RSG ratio Y/R, although small compared to the B/R ratio, {\it increases} with increasing $f$ ranging from: \e{2.6}{-2} at $f= 0.020$ to \e{5.1}{-2} at $f=0.031$. This small computed ratio is likely due to the reason that single star channel evolution may contribute relatively small fraction of YSGs that are found in nature compared to binary star evolutionary channels. However, we see that the number of YSGs and RSGs from the dataset are comparable (using the temperature range mentioned in Table \ref{tab:br_ratios} for YSGs and RSGs) in Fig. \ref{fig:13M_HR} and thus this ratio is close to unity.
}

\deleted{
In Fig. \ref{fig:13M_Teff_Yc} we show the evolution of  \Teff \ as a function of central helium abundance (which starts out at unity, since this is post TAMS evolution). At this metallicity ($Z=0.006$) the 13 \msun \ star makes a rapid transition across the  Hertzsprung Russell gap for all overshoot factors and spends its time in the RSG phase while it burns helium in its core. A few of them such as the green curve (with overshoot factor $f=0.020$) however make a single blue loop leading to higher \Teff \ at intermediate times before switching back to red again. The temperature excursion to smaller values as well as the starting point and "duration" in central helium fraction reduction depend upon the overshoot factors $f$. As noted already, beyond $f=0.31$, there is no further blueward excursion of the \Teff \ and these stars too (like the stars that undergo blue loops) explode finally as RSGs.
}

\deleted{
In contrast to the cases reported by\\ \cite{Ekstrom:2013aa} for 12 \msun \ and 15 \msun \ stars at SMC metallicity ($Z=0.002$) which enter the RSG stage at a more advanced stage of the core helium burning phase (see their Fig 1 (right panel)), all our models for 13 \msun \ enter the RSG phase at a very early stage of helium burning, and therefore our RSG phase is not as short as theirs in comparison to the respective BSG lifetimes. As a result, our B/R supergiant ratios are within a factor of two of the observed B/R in the LMC. 
}

\subsection{Convective overshoot and internal structure of the core connected to explodability} \label{sec:conv_core}

\deleted{The Kippenhahn diagram in Fig. \ref{fig:13M_kipp} shows the succession of different convective burning phases and zones inside the star during initial hydrogen-helium burning phases as well as more advanced phases of nuclear burning. The lifetime of stars from carbon burning onwards becomes much shorter than those at hydrogen or helium burning stages. This is due to higher core temperatures which leads to direct neutrino cooling dominating over energy transport due to radiation. Energy is also transported by convection inside the star.}

\deleted{The location of the lower boundary of carbon burning convective shells is an important factor \\ \citep{Sukhbold:2014aa} in determining the core bounce compactness factor %\footnote{$\xi_{M}$ is the progenitor's compactness parameter at core bounce - a dimensionless variable 
%%$\genfrac{M/\msun}{(R(M)/1000 km)}|_{t-bounce}$
%that allows a prediction of post-bounce dynamics and the possible evolution towards a black hole formation \citep{Oconnor:2011aa}. While the time of core bounce is critical for the evaluation of $\xi_{2.5}$ since this time is the only unambiguous time that gives the initial conditions for post-bounce evolution. nevertheless, $\xi_{2.5}$ has been utilized at slightly earlier times. %%by \citet{Oconnor:2011aa}. 
%It is thus instructive to see how the internal structure of the core represented by this parameter evolves as a star goes through the advanced stages of nuclear evolution in its core as in \citet{Sukhbold:2014aa}. The mass M = 2.5 \msun \ was chosen %%by \citet{Oconnor:2011aa} 
%since this is the mass scale for black hole formation. On the other hand \citet{Ugliano:2012aa,Ertl:2016aa} have suggested that a different fiducial mass 1.75 \msun \ may be a better discriminant of the explosion characteristics. As noted by \citet{Sukhbold:2018aa}, linked convective shells give a structure with smaller compactness parameter $\xi_M$ and thus favor SN explosions rather than BH formation.} 
during the core-collapse}
%
%\begin{equation}
%\xi_M = \frac{M/M_{\odot}}{R(M_{baryonic} = M)/1000 km}\Big |_{t-bounce}
%\end{equation}
%

The  \added{compactness} parameters $\xi_{2.5}$, $\xi_{1.8}$  at the presupernova stage and other advanced nuclear burning stages are illustrated in Fig. \ref{fig:13M_compactness} for different values of the overshoot parameter $f$. The asymptotic value of the $\xi_{2.5}$ at the presupernova stage clearly depend on the overshoot factor. \deleted{In general the larger the overshoot $f$, larger is the compactness parameter $\xi_{2.5}$ (and therefore harder it is to explode such a core) although the behavior near $f=0.032$ is not monotonically determined by $f$.} Note that these effects are due to the "non-burn" overshoot only as the overshoot in the metal (Z) burning stages are not included here. Nevertheless the effects on the compactness parameter at the various stages of core contraction and shell ignition are noticeable in the peaks and drops corresponding to the radial movement of the mass coordinate at 2.5 \msun. While the compactness evolution are relatively "quiescent" at smaller overshoot factors, the peaks and drops are substantial during the Ne, O, and Si burning stages for larger overshoot factors. The variation of the asymptotic behavior of  $\xi_{1.8}$ \added{at the end stage}\deleted{ (this factor relates to the compactness factor in the region just outside the neutron star boundary, if a neutron star is to finally form)} with respect to the overshoot factor $f$ is not congruent to the corresponding asymptotic behavior of  $\xi_{2.5}$.  The excursions for $\xi_{1.8}$ are larger (note the scales in the left part of Fig.  \ref{fig:13M_compactness}). This is because the region from where the neutron star would form sits in a deeper potential well than the part that is near the carbon core boundary characterized by $\xi_{2.5}$. \deleted{For larger overshoot $f$, there is a tendency of $\xi_{1.8}$ to rise substantially even at the late stages of Fe core contraction due to the change in lepton factor $Y_e$.}

%%%% Figure %%%%%%
%\begin{figure}[htb!]
%\plotone{Fig7.pdf}
%\caption{Evolution of M$_4$ (see subsection \ref{sec:conv_core} for definition) with time before CC for M$_{ZAMS}$= 13 \msun \  and Z = 0.006 models for different values of overshoot parameter $f$ with $f_0$ held constant at 0.005. 
%\label{fig:13M_M4}}
%\end{figure}
%
%\begin{figure}[htb!]
%\plotone{Fig8.pdf}
%\caption{Evolution of $\mu_4$ (see subsection \ref{sec:conv_core} for definition) with time before CC for M$_{ZAMS}$= 13 \msun \  and Z = 0.006 models for different values of overshoot parameter $f$ with $f_0$ is held constant at 0.005. 
%\label{fig:13M_mu4}}
%\end{figure}

\begin{figure}[htb!]
\includegraphics[width=0.48\textwidth]{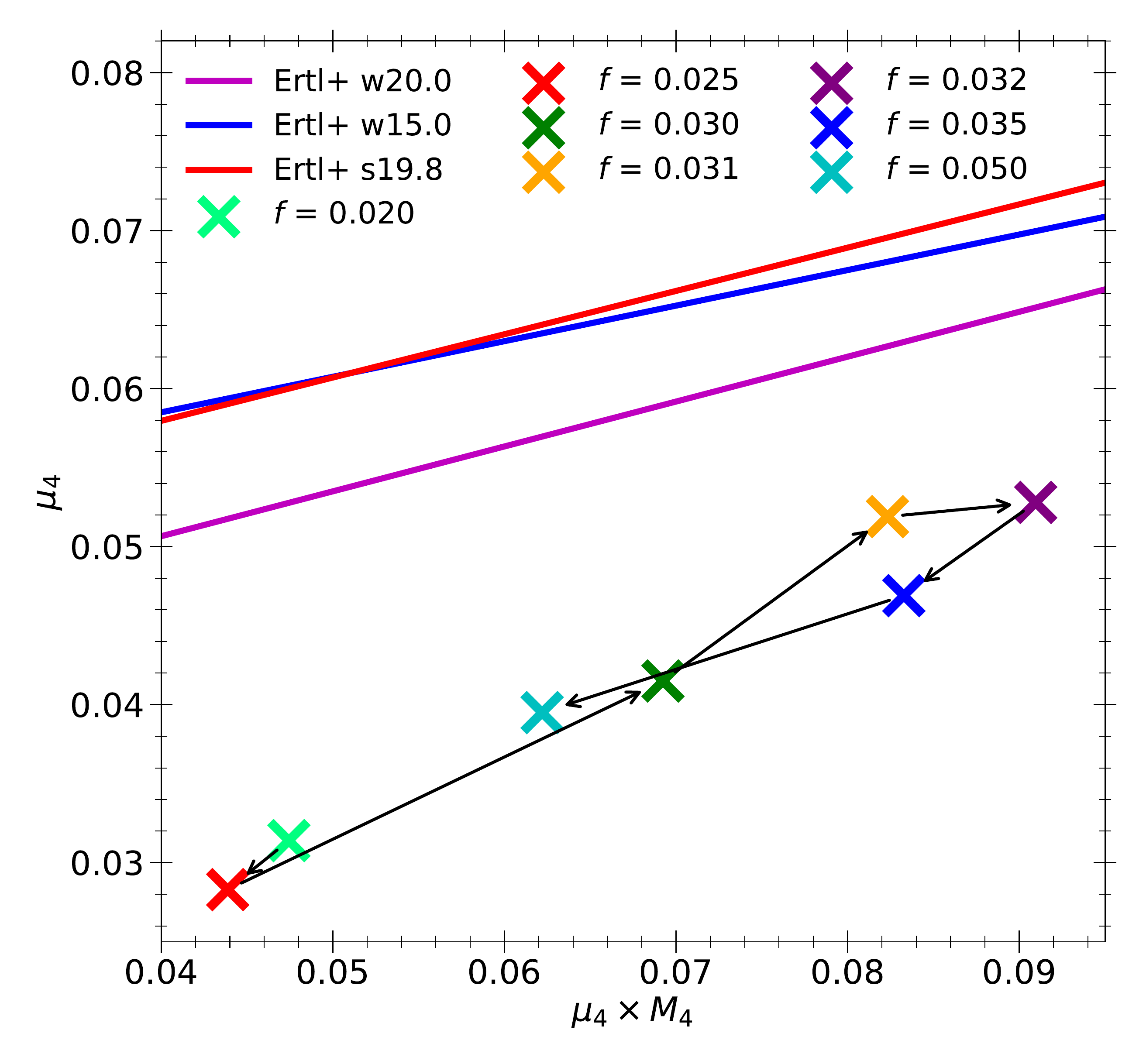}
\caption{A plot showing $\mu_4$ vs $mu_4 \times \rm M_4$ (see subsection \ref{sec:conv_core} for definitions) at the time of CC for M$_{ZAMS}$= 13 \msun \  and Z = 0.006 models for different values of overshoot parameter $f$ with $f_0$ is held constant at 0.005. Two straight lines with fit functions provided by \citet{Ertl:2016aa} for M$_{ZAMS}$= 15 \msun \ (BSG progenitor at CC) \& 18 \msun \ (RSG progenitor at CC) as separation line for calibration models are also plotted as examples. These lines separate the BH forming and SN-exploding stars.
\label{fig:13M_mu4M4}}
\end{figure}

\begin{figure*}[htb!]
\gridline{\fig{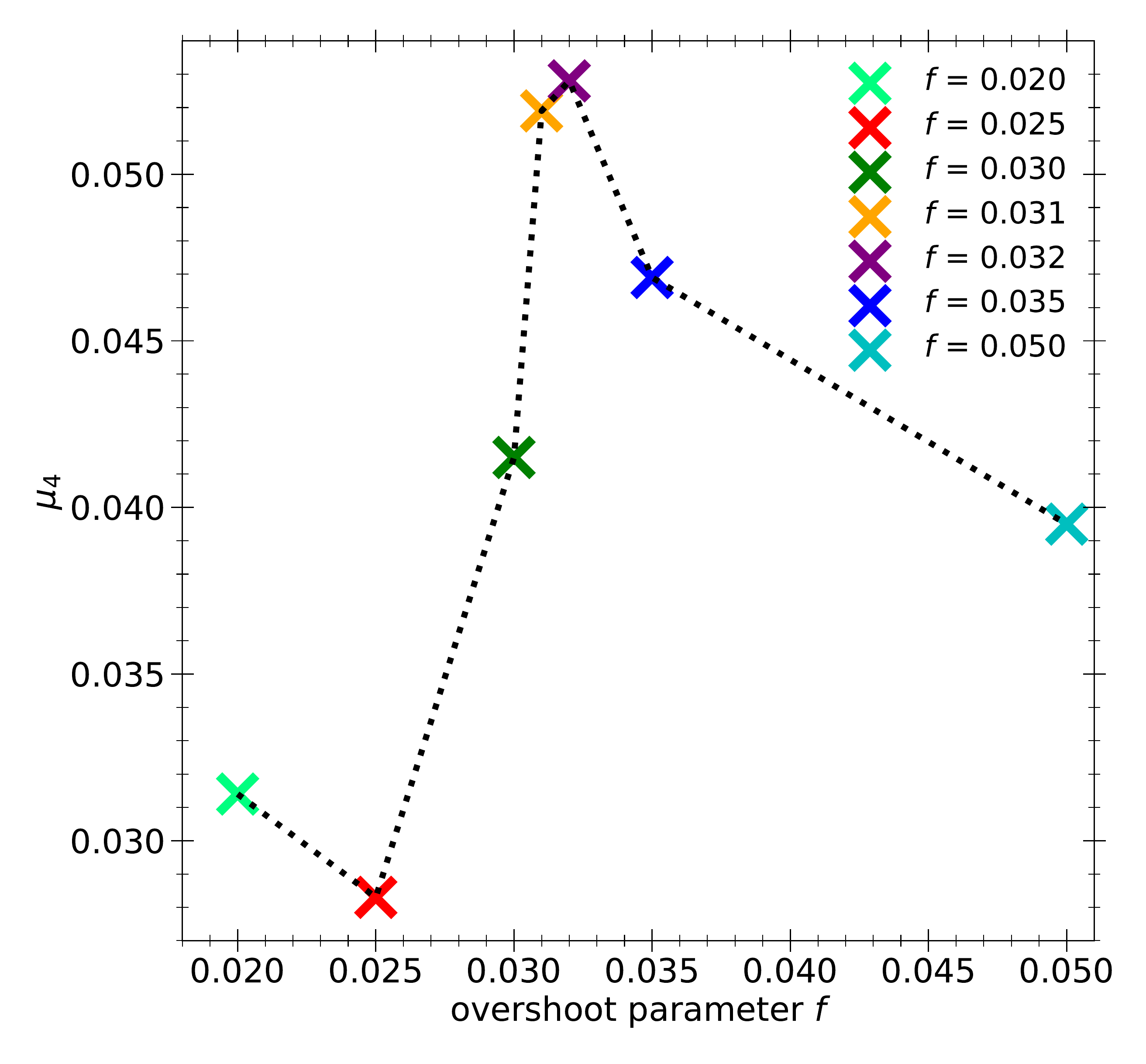}{0.45\textwidth}{(a)}
				\fig{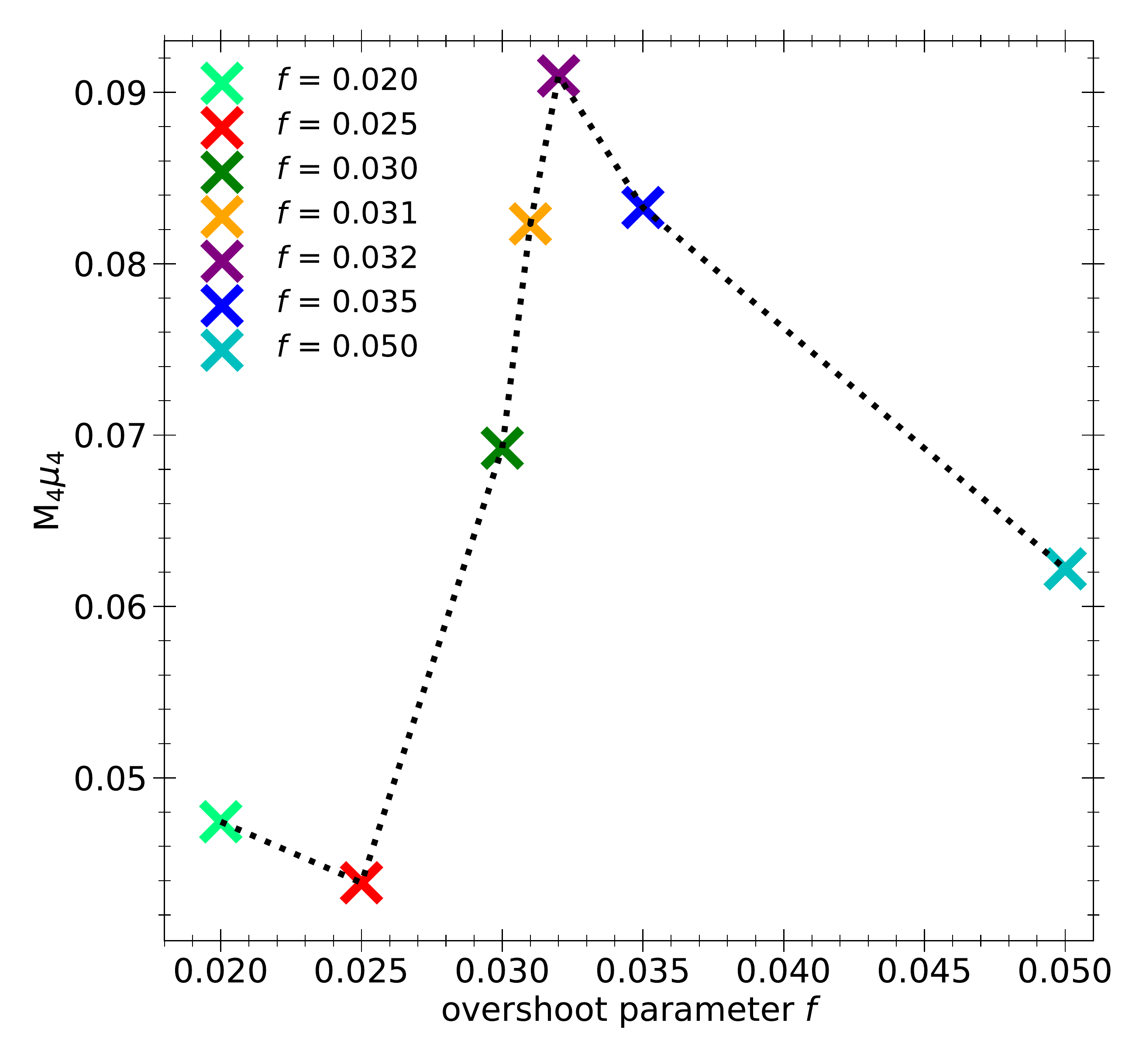}{0.45\textwidth}{(b)}}
\caption{\added{$\mu_4$ and $\mu_4 \times$ M$_4$ values at CC plotted as a function of f in panels(a) \& (b), respectively, for M$_{ZAMS}$= 13 \msun \  and Z = 0.006 models with different values of overshoot parameter $f$, where $f_0$ is held constant at 0.005.
\label{fig:13M_mu4_mu4M4_vs_f}}}
\end{figure*}
%%%%%%%%%%%%%%%%%

\deleted{
It is well known that the \added{success or failure of}\deleted{ explosion possibility versus collapse to a black-hole} is related to the strength and evolution of the accretion after stellar core-bounce \added{\citep[see e.g.][and references therein]{Ugliano:2012aa}}. While the expansion of the stalled shock of the initially exploding star is retarded by the ram pressure of the infalling stellar-core matter overlying the stalled shock, the shock expansion\footnote{\added{\citet{Colgate:1989aa} pointed out that as the inelastic collision behind the shock front convert the relative kinetic energy of moving matter to heat, they lead to the increase of the internal energy of the flow. This render the flow subsonic just behind the strong shock. \citet{Colgate:1971aa} showed that when the initial driving pressure due to the core elastic bounce subsumes, the resulting rarefaction wave soon able to catch up with the shock front, which then weakens. The presence of a hot, high-entropy bubble behind the shock due to deposition of neutrino energy in the hot bubble \citep{Mayle:1991aa,Wilson:1993aa} rescues the situation referred to above. The shock is ultimately re-invigorated only due to the presence of this hot, high-entropy bubble that keeps pushing against the matter outside the proto-NS (PNS) and makes the shock resumes its outward expansion. %\citet{1971ApJ...163..221C} showed that the matter adjacent to the proto neutron star immediately after the bounce cools so rapidly by neutrino emission that re-implosion occurs and a rarefaction wave catches up with the matter that was previously moving out with the bounce shock.
}} ("revival") is assisted by neutrino energy deposition behind the stalled shock \citep{Bethe:1985aa,Ertl:2016aa} beyond a critical neutrino luminosity \citep{Ray:1987aa,Burrows:1993aa}. At the same time the mass accretion rate on the proto neutron star, the source of the external ram pressure can be related to $\xi_{2.5}$ which is higher for denser stellar cores. \citet{Ertl:2016aa} then proposed a two parameter criterion which would in combination successfully predict the explosion behavior for neutrino driven supernova. These two parameters are $M_4$ and $\mu_4$ where $M_4$ is the enclosed mass (normalized to solar mass) where the dimensionless entropy per nucleon reaches $s = S/k_B = 4$ and 
%
%\begin{eqnarray} \label{eqn:mu4}
%\mu_4 & = & \frac{(\delta m/ M_{\odot})}{(\delta r/1000 km)}|_{s=4} \nonumber \\
%		   & = & \frac{(M_4 +\delta m/M_{\odot})-M_4}{[r(M_4 +\delta m/M_{\odot}) - r(s=4)]/1000 km }
%\end{eqnarray}
%
is the normalized mass derivative at this location (where $S/k_B =4$). Here $\delta$m = 0.3 \msun . \deleted{We use a slightly different definition of $\mu_4$ in the second part of equation \ref{eqn:mu4}, which is}
%
%\begin{equation}
%\mu_4 =  \frac{(M_4 +\delta m/2M_{\odot})-(M_4 -\delta m/2M_{\odot})}{[r(M_4 +\delta m/2M_{\odot}) - r(M_4 -\delta m/2M_{\odot})]/1000 km }
%\end{equation}
%
}

The parameters \added{$\mu_4$ and $\mu_4 \times$ M$_4$} are directly connected to the mass infall rate $\dot M$ on the proto-neutron star (PNS) and the electron neutrino luminosity ($L_{\nu_e}$) of the PNS\added{, respectively}. \added{As discussed earlier in section \ref{sec:background}, \citet{Ertl:2016aa} use these two parameters} to separate the black-hole forming and SN-exploding progenitor properties in the $M_4 \mu_4 - \mu_4$ plane. \added{They define a "separation line" using different calibration models that are compatible with the observations of SN 1987A}.

%During the time of oxygen burning to Si ignition there is a linear rise in the mass $M_4$ at the point where entropy is $s=4$ as a function of $log_{10}(\tau_{CC}-\tau)$ as seen in Fig. \ref{fig:13M_M4}, at least for the smaller values of the overshoot factor $f$. During this time, the material that will eventually end up collapsing to a neutron star moves out to larger core mass from about 1.5 \msun to about 1.7 \msun. For the largest overshoot factor, this development continues with larger excursions during the Si-burning and Fe-core contraction phases.
%In Fig. \ref{fig:13M_mu4} we show the variations of $\mu_4$ during various nuclear burning processes. Note that $\mu_4$ reaches asymptotically flat values during the Fe-core contraction phase immediately prior to core collapse. While at low overshoot factors $f$ the $\mu_4$ remains relatively low, at higher $f$ the asymptotic values of $\mu_4$ do not show any specific trend with $f$. 

In Fig. \ref{fig:13M_mu4M4}, we show the asymptotic values of $\mu_4$ plotted against the asymptotic values of  $\mu_4\times$M$_4$ at CC for our model simulations. The points in the figure span roughly a factor of two spread from the lowest to highest values. For models \deleted{which have overshoot factors}\added{with overshoot parameter values $f \le$ 0.031} that characterize blue loop behavior, increasing $f$-factors tend to span a growth \added{movement to the upper right} with an approximately positive slope in the 2-D plot. \deleted{while}\added{On the other hand,} for larger $f$-factors that characterize models that remain in RSG, increasing $f$ factors lead to a doubling up behavior with movement towards the lower left of this diagram. \deleted{The fitted straight lines separating the BH forming and SN-exploding progenitors  for two of the calibration models}\added{We have also shown a few separation lines} from \citet{Ertl:2016aa} \deleted{are also shown} in our figure. \deleted{Although, these two fits are calibrated with 15 \msun \ BSG progenitors at CC (as in the case of SN 1987A) and 18 \msun \ rotating RSG progenitor at CC, they give us a rough estimate for which of these models will be ``explodable".} \added{We note that these lines are calibrated for SN 1987A using BSG progenitor stars at the time of explosion. Unfortunately, we do not have calibrators like SN 1987A for CCSNe arising out of comparatively lower mass progenitors like SN 2013ej. Although \citeauthor{Ertl:2016aa} mention Crab as a low mass calibrator, critical information like neutrino luminosity, neutrino emission time-scale, and progenitor properties are missing for Crab. Thus, the separation lines shown here are only for a qualitative indication of which models might be "explodable". Points well above the separation lines (which depend upon the calibration model used) are expected to be failed explosions while those well below are likely to turn out to be successful SNe. The top panel of Fig. 11 of \citeauthor{Ertl:2016aa} shows a plot of these two parameters for a range ZAMS masses between 9--30 \msun . The range of $\mu_4\times$M$_4$ values in their Fig. 11 for models with ZAMS mass of about 13 \msun \  (by visual inspection) are in the range of 0.04--0.065 and that of $\mu_4$ are in the range of 0.03--0.053. Our values for these parameters are inclusive of their values. We note that for the 13 $M_{\odot}$ star considered here, all values of the overshoot factor lead to core properties that are far from the "separation line", and into the "exploding" region in the two dimensional parameter space. We therefore expect that collapse starting from these 1-d hydrodynamic stellar models are likely to produce explosions.} 

\added{In Fig. \ref{fig:13M_mu4_mu4M4_vs_f}, we show the variation of asymptotic values of $\mu_4$ and $\mu_4 \times$ M$_4$ at CC as a function of $f$. We can see that these quantities peak around f = 0.032 and fall off on both sides of this $f$ value. While both $\mu_4$ and $\mu_4 \times$ M$_4$ as a function of $f$ are high near the peak, the stellar structures for which both $\mu_4$ and $\mu_4 \times$ M$_4$ are relatively low are most favorable for explosion, as seen from Fig \ref{fig:13M_mu4M4}. Values of $\mu_4$ and $\mu_4 \times$ M$_4$ are relatively low at %both 
the low %and the high 
end of the $f$ values. %Thus, these values are more favorable for explosion because they are more distant from the line of separation of exploding versus non-exploding models.
} 

\section{Discussion and Conclusion}\label{sec:conclusion}

\deleted{The blue vs RSG ratios are known to depend upon the metallicity of the environment in which they are found (see e.g. \citet{Meynet:2011aa}).} \added{We show results for simulations of a 13 \msun \ progenitor in light of SN 2013ej using }\deleted{ The metallicity for which we show our results here have been chosen as $Z= 0.006$}\added{metallicity of $Z= 0.006$ which is typical of the host galaxy M74 for the SN. We systematically explore the convective overshoot parameter $f$ between a value of 0.010 and 0.05 to study the effects on the internal structure and the surface properties of the star}. \deleted{This}\added{Our value of metallicity} is very close to the LMC metallicity of $Z=0.007$ \citep{Evans:2008aa}. \deleted{They}\added{\citeauthor{Evans:2008aa}} also note that the mean mass of the LMC stars in their sample is 13 \msun , which is the ZAMS mass for which we present our results here.

We find that the appearance of the blue loop in the HRD is dependent on the extent of overshoot mixing. For the range of $f$ values studied here, the blue loops only appear for \added{0.020 $\le$} $f \le$ 0.031. The \added{lower} or higher $f$ values \added{do not lead to any blue loop}. We attribute the blue loops to the presence of fully radiative hydrogen rich envelope (``non-burn" envelope, see Fig. \ref{fig:13M_kipp}) during the core He-burning stage in the low $f$ simulations as opposed to the high $f$ simulations (see section \ref{sec:results} for details). The appearance of the blue loop is not affected by the choice of the other overshoot parameter $f_0$. The higher $f$ simulations also end up in a slightly lower mass pre-SN star (by $\sim$0.5 \msun, \added{see Fig. \ref{fig:13M_kipp}}). These model stars spend more time in the RSG phase, during which the mass loss rate is higher compared to the BSG phase\added{, and they end up with lower pre-SN masses.}

The surface CNO abundance evolution also indicates a connection to the appearance of blue loop in the HRD. The lower $f$ simulations show higher enrichment soon after core helium ignition compared to the higher $f$ simulations (see Fig. \ref{fig:13M_CNO}) \deleted{due to a rapid dredge-up of the CNO nuclei and less efficient mixing in the envelope for a shorter duration as the convection shuts off soon after in the low $f$ simulations}. Another episode of surface CNO enrichment occurs in the low $f$ simulations \added{($f \le$ 0.031 which characterize the blue loops)} when helium is almost depleted at the center, when the convective envelope again penetrates deeper into the star \added{(see Fig. \ref{fig:13M_kipp})}. At this point of evolution, in the higher $f$ simulations \added{($f \ge$ 0.031)}, this enrichment is marginal at best. 

\added{The current observational data using VLT-FLAMES \citep{Hunter:2007aa} is lacking the precision and suffers from large uncertainties. Hence, these data cannot distinguish between the surface CNO abundances at different stages of stellar evolution that our models predict. Some of these observed stars are spectroscopic binaries or Be stars, and are richer in surface nitrogen \citep[ref.][]{Maeder:2014aa} and are not appropriate stars to compare. Similarly, the ratios of surface nitrogen to carbon abundances at different stages (see Fig. \ref{fig:13M_NC_NO_vs_f}, panel (a)) are indistinguishable in observations,  due to large uncertainties in both surface nitrogen and carbon abundances. However, the ratios of surface nitrogen to oxygen abundances can marginally distinguish between different evolutionary stages in the best case scenario (i.e. in case of the smallest observed uncertainties), but the differences in surface N/O ratios (see Fig. \ref{fig:13M_NC_NO_vs_f}, panel (b)) for different $f$ values at the late evolutionary stages that our models predict are indistinguishable even for the best case.}

\added{\citet{Davies:2019aa} present a method of inferring the mass of the progenitor of a type IIP SN using the observed abundance ratios at a very early stage after the explosion of the SN. During this early phase (within a few days) while the photosphere is very hot, optical spectra show ionized carbon, nitrogen and oxygen as also hydrogen and helium. They argue that predictions from stellar evolution calculations for RSGs for the terminal surface [C/N] ratio is correlated with the initial mass of the progenitor star. The use of very early spectra of the supernova according to them facilitates the estimation of the pre-explosion carbon, nitrogen, and oxygen abundances, since the photosphere being hot, high ionization species of such elements allow the spectra to be dominated by these elements. While very early time spectra of supernovae have been somewhat rare so far (e.g. SN 2013fs = iPTF13dqy), where O \textsc{vi}, O \textsc{v}, O \textsc{iv} and N \textsc{v} lines have been detected at $T_{in} = 48-58 kK$ \citep{Yaron:2017aa} or SN 2016esw \citep[typically at 0.4-0.6 d after explosion,][]{de-Jaeger:2018aa} modeling such lines as O \textsc{v} and O \textsc{vi} have not been very reliable. Moreover, very early spectra are often heavily dominated by high continuum emission and line characteristics may be more difficult to determine.}

In our HRD (Fig. \ref{fig:13M_HR}), we have plotted supergiants from various archival data sets overlapping the HR tracks from our simulations. We find a good number of observed supergiants in LMC  overlapping the evolutionary tracks\added{, both in the blue and the red parts of the HRD}. With the high-precision observations of g-modes \citep[see e.g.][]{Georgy:2014aa, Saio:2013aa}, one may be able to distinguish between the stars that are either massive stars with higher luminosity transitioning to RSG stage for the first time or the less massive stars going through the blue loop.

\deleted{\citet{Sukhbold:2014aa, Sukhbold:2018aa} have explored presupernova core structures of massive stars, especially the core compactness parameters $\xi_{2.5}$, $\xi_{1.75}$, 
etc., for a large variety of initial stellar masses, metallicities and other physical and code inputs. Many characteristics of presupernova stars, e.g. He, C, Fe core masses, their luminosities and radii varied smoothly with initial mass. However, in the range of masses near 14 -19 \msun, they found that the compactness parameter vary wildly with mass in a non-monotonic manner and it was found to be sensitive to events in the last few years of the star's life. They also found significant differences in the compactness parameters for stars with the same mass but with different metallicities. They ascribed the non-monotonic variation of the compactness parameters to the interaction of convective carbon and oxygen burning shells inside the star. Variations in the location of oxygen burning shells were in turn influenced by the locations of the previous history and location of carbon burning shells. As the relatively lighter (among massive) stars go through their advanced stages of nuclear evolution, they can have three or more carbon burning shells over the carbon burning core and their central regions become increasingly degenerate soon after the carbon depletion at the stellar center. The shells burning outside the Chandrasekhar mass but inside the mass coordinate 2.5 \msun \ end up changing the compactness parameter.} In the present paper, we study both internal and surface properties of the star to different treatments of convection in terms of overshoot factors for a fixed initial mass and at a fixed metallicity. We demonstrate in Paper II, that our models are computed with adequately fine zoning to resolve the density and pressure gradients. We report in this paper the strong variation of the shell boundaries influencing the compactness parameter depending upon the convective overshoot factors. \deleted{While at low overshoot factors the influence on the pre-SN compactness factors may be monotonically increasing, at slightly higher overshoot factors, the compactness factors  do not increase monotonically. This is demonstrated in Figs. \ref{fig:13M_compactness} and \ref{fig:13M_mu4M4} above.}
\added{The compactness parameters $\xi_{2.5}$ resulting from the overshoot parameter range ($f$ = 0.02 to 0.030) varies by a factor of 2.5, but thereafter for higher values of f, has non-monotonic variation. The parameters $\mu_4$ and $\mu_4 \times$M$_4$ have an approximately bell shaped variation with $f$, peaking at $f$= 0.032. This is demonstrated in Figs. \ref{fig:13M_compactness} and \ref{fig:13M_mu4_mu4M4_vs_f} above. Our results predict that all simulations of approximately 13 \msun \ star with overshoot factors considered here would most likely explode in to a SN, based on the criteria suggested by \citet{Ertl:2016aa}}

\section{Acknowledgments}
We thank the directors and the staff of the Tata Institute of Fundamental Research (TIFR) and the Homi Bhabha Center for Science Education (HBCSE-TIFR) for access to their computational resources. This research was supported by a Raja Ramanna Fellowship of the Department of Atomic Energy (DAE), Govt. of India to Alak Ray and a DAE postdoctoral research associateship to Gururaj Wagle. The authors thank the anonymous referee for his/her constructive comments that helped us to improve this paper. %GW thanks Rob Farmer for valuable feedback through private communication. 
Ajay Dev and Adarsh Raghu thank the NIUS program at HBCSE (TIFR).
The authors acknowledge the use of NASA's Astrophysics Data System and the VizieR catalog access tool, CDS, Strasbourg, France.

\software{MESA r-10398 \citep{Paxton:2011aa,Paxton:2013aa,Paxton:2015aa,Paxton:2018aa}, Anaconda Spyder (Python 3.6)} 

\bibliography{my_bib}
%%%%%%%%%%%%%%%%%%%%

\end{document}